\documentclass[draft]{agujournal2019}
\usepackage{url} 
\usepackage{lineno}
\usepackage{caption}
\usepackage{soul}
\usepackage{amsmath}
\usepackage{tikz}
\usepackage{subcaption} 
\usepackage{booktabs} 
\usepackage{float}
\usepackage{amssymb}

\draftfalse

\journalname{Journal of Advances in Modeling Earth Systems}

\begin{document}

\title{FAST-ML: A Hybrid Physics—Machine Learning Framework for Tropical Cyclone Intensity Forecasting}

\authors{Shijie Xiao\affil{1,4}, Jonathan Lin\affil{2}, Thomas Ehrmann\affil{3}, and Ali Sarhadi\affil{1,4}}

\affiliation{1}{School of Earth and Atmospheric Sciences, Georgia Institute of Technology, Atlanta, GA, USA}
\affiliation{2}{Department of Earth and Atmospheric Sciences, Cornell University, Ithaca, NY, USA}
\affiliation{3}{Sandia National Laboratories, Albuquerque, NM, USA}
\affiliation{4}{School of Interactive Computing, Georgia Institute of Technology, Atlanta, GA, USA}

\correspondingauthor{Shijie Xiao}{sjxiao@gatech.edu}

\begin{keypoints}

\item A differentiable reduced-order hurricane model learns unresolved environmental ventilation closures directly from observations.

\item Observation-driven closure learning improves tropical cyclone intensity forecasts and nearly halves rapid-intensification false alarms.

\item Learned closures are consistent with ventilation theory and support interpretable, computationally efficient intensity forecasting.

\end{keypoints}

\begin{abstract}

Rapid intensification (RI) remains one of the most consequential and difficult aspects of tropical cyclone (TC) forecasting. Although full-physics numerical weather prediction models can represent the processes governing RI, resolving storm-environment interactions remains computationally expensive, while purely data-driven approaches often lack physical interpretability. We present FAST-ML, a hybrid framework that bridges data-driven efficiency with physical constraints. A physically informed dual-stream neural parameterization ingests 3D ERA5 fields to diagnose ventilation controls---environmental wind shear and mid-level entropy deficit. By optimizing these parameters end-to-end through a differentiable FAST intensity model, this architecture establishes a robust new paradigm for observation-driven parameter optimization, ensuring storm evolution remains strictly governed by thermodynamic principles. By better capturing the storm's continuous intensity evolution, FAST-ML improves upon its physical baseline, reducing ensemble CRPS across forecast lead times, with a reduction of approximately 31\% at 60 h and nearly halving the RI false alarm ratio without sacrificing detection skill. In a 100-member ensemble configuration, FAST-ML produces intensity forecasts comparable to FNV3 for selected storms under the evaluated input configurations. Furthermore, zero-shot tests on selected Eastern Pacific storms provide encouraging evidence of cross-basin transferability. FAST-ML provides a modular intensity forecasting framework that can be coupled with externally supplied storm tracks and environmental fields. It demonstrates that observation-driven parameter learning within physically constrained dynamics simultaneously enhances accuracy, interpretability, and computational efficiency.

\end{abstract}

\section*{Plain Language Summary}
Tropical cyclones can strengthen very rapidly, sometimes gaining more than 30 knots of wind speed in one day. Predicting these RI events remains a major forecasting challenge. Detailed physics-based weather models capture intensification processes but are computationally expensive, limiting repeated runs for uncertainty estimation. Artificial intelligence (AI) forecasts are faster but often offer limited insight into their physical reasoning. We developed FAST-ML, combining physics and AI. Rather than directly predicting intensity, AI analyzes surrounding winds, temperature, and humidity to estimate key environmental factors controlling storm growth. These drive a physics-based hurricane model governing storm evolution, preserving physical consistency while learning complex environmental relationships from data. The AI learned environmental relationships consistent with established understanding of storm thermodynamics without these relationships being explicitly prescribed. On Atlantic hurricanes, FAST-ML improved intensity forecasts over the original physical model, substantially reduced RI false alarms, and generated probabilistic forecasts competitive with a leading AI weather model using far fewer computational resources. Strong performance on Pacific storms excluded from training indicated transferable physical relationships rather than memorization of historical patterns. These results show that combining AI with physical principles can improve hurricane intensity forecasting while maintaining interpretability, efficiency, and scientific credibility.

\newpage

\section{Introduction}
TCs are among the most damaging weather-related hazards worldwide, producing extreme winds, heavy rainfall, storm surge, and inland flooding that collectively lead to substantial economic losses and societal disruption each year. While major advances in satellite observations, aircraft reconnaissance, data assimilation, and numerical weather prediction (NWP) have significantly improved track forecasts over recent decades, progress in intensity forecasting has been comparatively slower. Forecast errors associated with intensity changes remain a central limitation of operational hurricane prediction, particularly during episodes of RI, when maximum sustained winds can increase by more than 30 kt within 24 hr. Such events frequently occur near landfall and can substantially reduce preparation time, making RI one of the most societally consequential challenges in tropical meteorology. \cite{kaplan2010, emanuel2017}.

The difficulty of TC intensity forecasting originates from the complex, non-linear nature of TC evolution. While storm intensity is driven by multiscale interactions between internal vortex dynamics and environmental factors, forecasting, particularly for RI, is fundamentally constrained by the system's extreme sensitivity to initial states. For instance, RI evolution is highly sensitive to the initial inner-core moisture profile \cite{riemer2010, emanuel2012}. Furthermore, the underlying thermodynamic framework is strongly regulated by environmental ventilation, which exhibits pronounced non-linearities characterized by intensity bifurcation points \cite{Tang2012}. Depending on the initial intensity and the magnitude of ventilation forcing, the system may rapidly diverge toward entirely different equilibrium solutions, either successfully closing the thermodynamic cycle to undergo RI or decaying due to the intrusion of low-entropy air \cite{Tang2010, Tang2012,xu2023}. Consequently, accurate intensity forecasting requires not merely representing the available thermodynamic potential, but also precisely resolving the highly sensitive ventilation parameterizations that dictate which evolutionary pathway a storm will follow.

Traditional intensity forecasting primarily relies on high-resolution numerical weather prediction systems. By solving the governing equations of atmospheric motion and thermodynamics, NWP models preserve physical consistency while providing explicit representations of storm structure and environmental forcing. However, reliable prediction of intensity change, and particularly RI, often requires resolving inner-core processes such as eyewall convection, vortex asymmetries, moist entropy transport, and air--sea coupling. Accurately capturing these processes demands increasingly refined spatial resolution and sophisticated physical parameterizations, resulting in substantial computational costs. Although operational forecasting systems continue to improve, producing the large ensembles needed for uncertainty quantification remains computationally expensive, especially when forecasts must be updated in real time \cite{judt2019, LinFHLO}.
This computational challenge has motivated considerable interest in reduced-order dynamical models \cite{Dem09} that retain the dominant thermodynamic controls on storm intensity while dramatically reducing computational complexity. Such models provide an attractive alternative for large-ensemble forecasting because they explicitly represent the dominant physical mechanisms governing intensity evolution while evaluating thousands of plausible trajectories at only a fraction of the cost of full-physics NWP simulations, making them particularly suitable for computationally efficient and physically consistent probabilistic forecasting and risk assessment applications.
Among these reduced-order approaches, the FAST TC intensity model \cite{emanuel2017} represents a particularly promising framework. FAST describes storm evolution through a physically based system of ordinary differential equations that interprets the TC as an idealized heat engine embedded within a time-varying environment \cite{emanuel1986air,Ema95}. The model explicitly incorporates physically meaningful controls including potential intensity, ocean feedbacks, environmental entropy deficits, and vertical wind shear, providing a computationally efficient approximation to more comprehensive dynamical simulations. The resulting framework has demonstrated considerable utility for probabilistic forecasting, synthetic storm generation, and large-ensemble hurricane risk analysis \cite{lin2023, LinFHLO, emanuel2017,sarhadi2024climate}.

Despite its strong physical foundation, the predictive skill of FAST—and reduced-order intensity models more broadly—is fundamentally constrained by how unresolved environmental interactions are parameterized, particularly those associated with ventilation processes and storm–environment coupling. Because these processes cannot be resolved directly within reduced-order frameworks, their influence is commonly represented through simplified analytical formulations and empirically calibrated coefficients. While physically motivated, such parameterizations inevitably assume a reduced representation of the environment and may struggle to adapt to rapidly evolving storm-specific conditions.

Recent advances in machine learning provide a natural opportunity to address this limitation. Machine Learning and neural networks have demonstrated an exceptional ability to extract high-dimensional nonlinear relationships from large observational and reanalysis datasets and have rapidly emerged as powerful tools for weather prediction. Models such as FourCastNet, Pangu-Weather, GraphCast, FuXi, and GenCast have shown that data-driven approaches can successfully learn complex atmospheric evolution patterns while achieving unprecedented computational efficiency \cite{pathak2022, bi2023, lam2023learning, chen2023, price2025}.
However, the central challenge for TC intensity forecasting differs fundamentally from large-scale weather prediction. The objective is not merely to predict atmospheric states but rather to accurately represent the physical mechanisms that regulate vortex intensification and decay. While specialized data-driven models have been developed to predict TC RI directly \cite{Gri22, Che23}, purely data-driven models often learn effective statistical mappings between inputs and outputs but do not explicitly expose the governing thermodynamic pathways responsible for intensity change \cite{emanuel1999, kashinath2021,smithThorpe2026,gomez2026tcbench}. As a result, although they can provide highly competitive forecasts, they offer limited insight into the environmental controls driving individual storm evolution and can be difficult to interpret in a physically meaningful framework.

These considerations suggest that the most promising path forward may not be replacing physical models with machine learning, but rather embedding machine learning within physical frameworks to improve representations of unresolved processes while retaining physically constrained system evolution. This philosophy has motivated a rapidly growing class of physics--machine learning hybrid frameworks in Earth system science, in which data-driven methods are used to learn unresolved physical processes while the governing system dynamics remain constrained by established physical principles \cite{raissi2019physics,karniadakis2021, willard2022,smithThorpe2026}.
This paradigm has inspired hybrid approaches to TC intensity forecasting
that couple neural networks with physically based dynamical models \cite{Zho24}. For TC intensity forecasting, this perspective is particularly compelling because the dominant uncertainties are often associated with unresolved environmental interactions rather than deficiencies in the governing thermodynamic framework itself. Consequently, instead of learning storm intensity directly, machine learning can be used to infer the latent environmental quantities and state-dependent closures that regulate the physical model. Such a formulation preserves scientific interpretability, maintains thermodynamic consistency, and avoids replacing well-established physical relationships that are already known to be effective \cite{schneider2017,rasp2018}, which is critical for physically trustworthy future projections. This distinction is particularly crucial in the context of a non-stationary climate. Purely data-driven models, which excel at interpolating within historical distributions, often struggle with out-of-distribution generalization and can produce physically implausible predictions under unprecedented future warming scenarios \cite{kashinath2021, beucler2021enforcing}. By confining the machine learning component to unresolved physical closures while governing the macro-state with strict thermodynamic bounds, our hybrid approach ensures that the model respects fundamental conservation laws and energy limits. This helps ensure that long-term intensity projections remain both dynamically consistent and physically reliable, even when extrapolating into unobserved climatic conditions.

Motivated by this concept, we develop a physics-based model with embedded machine learning for TC intensity forecasting. Unlike approaches that directly predict storm intensity from environmental fields, the proposed framework — which combines the FAST intensity simulator with a neural diagnostic module — uses machine learning exclusively to infer latent environmental closure parameters (e.g., ventilation and shear) from three-dimensional atmospheric and oceanic fields, while intensity evolution remains governed by a differentiable physical model. Rather than generating intensity forecasts independently, the neural network functions as a dynamic closure that supplies physically meaningful environmental forcings to a differentiable physical solver. Storm evolution remains governed by thermodynamic principles embedded within FAST, while unresolved environmental influences are inferred from high-dimensional observations.
This architecture establishes a continuous bridge between data-driven learning and process-based dynamics. The machine learning component provides adaptive state-dependent environmental parameterizations, whereas the physical model supplies mechanistic constraints and physically interpretable system evolution. Together, these complementary elements enable computationally efficient forecasting while preserving physical consistency across the full storm lifecycle.
We evaluate the framework using North Atlantic TCs and investigate its performance in both deterministic and probabilistic forecasting configurations. To further validate the model's structural design and cross-basin generalization, we incorporate an architectural ablation study and a zero-shot blind test on Eastern Pacific (EP) storms. Results demonstrate substantial improvements in intensity prediction accuracy, enhanced representation of RI events, and competitive probabilistic forecast skill. Furthermore, the learned environmental parameterizations recover established theories of mid-level ventilation and wind-shear regulation, achieving physical interpretability with a simpler and more computationally efficient architecture than purely data-driven alternatives \cite{Tang2010,Tang2012}. Overall, this study provides a general blueprint for embedding machine learning within reduced-order physical models, improving forecast skill while preserving physical consistency and interpretability, and enabling affordable large-ensemble forecasting for TCs and other Earth-system applications.

\section{Methodology}

\subsection{Physical Foundation: The FAST Intensity Model}
\label{FAST}

The proposed framework is built upon the Fast and Accurate Simulation of TCs (FAST) model \cite{emanuel2017}, a reduced-order dynamical framework that represents a TC as an idealized axisymmetric heat engine embedded within a time-varying environment. FAST provides an attractive foundation for this study because it explicitly represents the dominant thermodynamic controls on TC intensity evolution while remaining computationally efficient enough for large-ensemble probabilistic forecasting applications. This balance between physical realism and computational efficiency makes FAST particularly suitable for investigating whether machine learning can improve unresolved environmental parameterizations without replacing the governing physical dynamics.
The model evolves two prognostic state variables: the maximum azimuthal surface wind speed, $v$, and a nondimensional inner-core moisture variable, $m$, bounded between 0 and 1. Their temporal evolution is governed by \cite{emanuel2017}:

\begin{align}
\frac{dv}{dt}
&=
\frac{1}{2}\frac{C_k}{h}
\left[
\alpha \beta V_p^2 m^3
-
\left(1-\gamma m^3\right)v^2
\right],
\label{eq:dvdt}
\\
\frac{dm}{dt}
&=
\frac{1}{2}\frac{C_k}{h}
\left[
(1-m)v
-
\chi Sm
\right],
\label{eq:dmdt}
\end{align}

where $h$ is the atmospheric boundary-layer depth and $C_k$ is the surface enthalpy exchange coefficient, $V_p$ denotes the environmental potential intensity and represents the theoretical thermodynamic upper bound of the cyclone heat engine \cite{bister2002low}. The first equation governs the evolution of storm intensity through the competition between thermodynamic energy input and frictional dissipation. The second equation governs the evolution of inner-core moisture and contains the ventilation term $\chi Sm$, which acts as the primary mechanism through which the large-scale environment suppresses storm development.
All the rest thermodynamic coefficients are computed analytically from environmental conditions:

\begin{align}
\beta &= 1-\epsilon-\kappa,
\\
\gamma &= \epsilon+\alpha\kappa,
\\
\epsilon &= \frac{T_s-T_o}{T_s},
\\
\kappa &= \frac{\epsilon}{2}
\frac{C_k}{C_d}
\frac{L_v}{R_d}
\frac{q_s^*}{T_s},
\\
\alpha &= 1-0.87\exp^{-\tilde{z}},
\\
\tilde{z} &= 0.01\Gamma^{-0.4}h_m u_T V_pv^{-1}
\end{align}

The potential intensity, $V_p$, is computed as

\begin{equation}
V_p^2
=
S_w^2
\frac{C_k}{C_d}
\frac{T_s}{T_o}
\left(CAPE^* - CAPE\right),
\label{eq:potential_intensity}
\end{equation}

where $S_w$ is the surface wind reduction factor, $C_k$ and $C_d$ are the surface exchange coefficients for enthalpy and momentum, respectively, $T_s$ and $T_o$ denote the sea-surface and outflow temperatures, and $CAPE^*$ and $CAPE$ denote the convective available potential energy of saturated air at the sea surface and boundary-layer air, respectively. The remaining thermodynamic and ocean-coupling parameters include $q_s^*$, the saturation specific humidity at the ocean surface; $L_v$, the latent heat of vaporization; $R_d$, the gas constant for dry air; $\epsilon$, the thermodynamic efficiency; $\alpha$, the ocean interaction parameter; $\Gamma$, the sub-mixed-layer thermal stratification in $\text{K}\,(100\,\text{m})^{-1}$; $h_m$, the mixed-layer depth; and $u_T$, the translation speed. The parameter $\alpha$ accounts for ocean feedback processes through wind-induced upper-ocean cooling and depends on storm translation speed and upper-ocean stratification.

Although potential intensity ($V_p$) and ocean coupling ($\alpha$) determine the thermodynamic limits of the system, the realized intensity trajectory is strongly regulated by the environmental ventilation term, $\chi S$, in Eq.~\ref{eq:dmdt}. This term mathematically encapsulates the mid-level ventilation hypothesis \cite{Tang2010}, which dictates that the ventilation of low-entropy environmental air into the TC inner core is the primary pathway for vertical wind shear to weaken a storm. The FAST model elegantly decouples this suppressive mechanism into a wind shear driver ($S$) and a thermodynamic sink ($\chi$). The wind shear represents the magnitude of the environmental bulk vertical wind shear, typically evaluated between the 850 hPa and 250 hPa pressure levels. Physically, environmental shear tilts the TC vortex, generating asymmetric processes and exciting mesoscale eddies that act as a mechanical transport system \cite{riemer2010, All21}, forcing the mixing of environmental air into the protected inner-core eyewall \cite{riemer2010,Tang2010}. In contrast, the thermodynamic sink represents the environmental saturation entropy deficit. Following Tang and Emanuel \citeyear{Tang2012}, the generalized formulation for $\chi$ is defined as the mid-level entropy deficit normalized by the air-sea thermodynamic disequilibrium:

\begin{equation}
\chi = \frac{s_m^* - s_{env}}{s_s^* - s_b}
\end{equation}

where $s_m^*$ and $s_{env}$ denote the saturation and environmental pseudo adiabatic entropies at mid-levels (typically maximized between 500 hPa and 700 hPa), while $s_s^*$ and $s_b$ represent the saturation entropy at the sea surface and the entropy of the boundary layer. 

Crucially, the non-linear coupling between intensity and moisture in this framework mathematically dictates the storm's sensitivity to such environmental forcing. Setting the time derivatives in the governing equations to zero reduces the system to a steady-state cubic polynomial for intensity $v$:
\begin{equation}
(1 - \gamma)v^3 + 3(\chi S)v^2 - [\alpha\beta V_p^2 - 3(\chi S)^2]v + (\chi S)^3 = 0
\label{eq:cubic}
\end{equation}
This cubic formulation analytically demonstrates the existence of multiple equilibrium solutions and intensity bifurcation points dictated by the magnitude of the ventilation sink ($\chi S$). Consequently, directly inverting this dynamical system from sparse observations is highly ill-posed, as gradient-based optimization can easily converge to physically spurious branches. Our FAST-ML framework circumvents this equifinality by anchoring the optimization with robustly inverted initial states and physically consistent target outcomes (e.g., RI events). By optimizing the trajectory between known boundary conditions rather than isolating instantaneous states, the machine learning model successfully constrains the learned ventilation parameters to the appropriate solution branch, yielding physically consistent closures that match observations.

Among all terms appearing in Eqs.~(\ref{eq:dvdt})--(\ref{eq:dmdt}), the synergistic ventilation product $\chi S$ represents the largest source of structural uncertainty. In the original FAST formulation, both $S$ and $\chi$ are estimated using analytical relationships derived from coarse-resolution environmental diagnostics. Although physically motivated, such formulations necessarily simplify the spatial and vertical structure of the surrounding atmosphere through prescribed pressure levels, uniform radial mixing assumptions, and empirically calibrated coefficients. Consequently, they inadequately represent storm-specific interactions associated with asymmetric convection, localized humidity anomalies, and rapidly evolving shear environments. FAST-ML addresses this limitation by replacing the analytical ventilation closures with a neural diagnostic model that estimates the tendencies of $S$ and $\chi$ directly from three-dimensional environmental fields while preserving the original FAST governing physical equations.

To independently evaluate whether the neural closure recovers true thermodynamic mechanisms rather than fitting spurious statistical noise, we establish a physical verification benchmark grounded in the theoretical framework of Emanuel and Tang \cite{emanuel1988, Tang2010}. Fundamentally, environmental ventilation is driven by the influx of low-entropy air into the storm's inner core, primarily governed by the mid-level non-dimensional entropy deficit ($\chi$) and environmental vertical wind shear ($S$). Following Tang and Emanuel \cite{Tang2010, Tang2012}, the combined physical ventilation index ($\Lambda$) is expressed as:
\begin{equation}
\Lambda = \frac{S \chi}{V_{p}}
\label{eq:physical_ventilation_index}
\end{equation}
While the full index $\Lambda$ is non-dimensionalized by potential intensity ($V_p$), its numerator, $\chi S$, dictates the absolute magnitude of the environmental thermodynamic forcing. In our diagnostic evaluation, the actual environmental ventilation ($F_{\text{vent}}$) serves as the explicit physical benchmark for this absolute thermodynamic suppression, computed as the integral of eddy entropy flux across the inner-core cylindrical boundary ($r \approx 200\text{~km}$):
\begin{equation}
F_{\text{vent}} = \int_{z_0}^{z_{\text{top}}} \int_{0}^{2\pi} \left( \rho u' s' \right) r \, d\theta \, dz
\label{eq:eddy_entropy_flux}
\end{equation}
where $\rho$ is the air density, $u'$ is the asymmetric radial wind velocity, $s'$ is the asymmetric pseudo-adiabatic entropy anomaly (consistent with the thermodynamic entropy states $s$ defined previously), and $z$ represents the geopotential height (consistent with the atmospheric predictor variables), integrating vertically from the surface to the tropopause. Furthermore, to rigorously validate the physical realism of the learned parameterization, we evaluate the physical correlation between the net ventilation variable predicted by the neural network---derived explicitly as the multiplicative product $\chi S$---and this diagnostic environmental ventilation $F_{\text{vent}}$. This establishes a direct, verifiable link between the learned neural closure and established atmospheric thermodynamics.

To render the governing equations operational with best-track observations, the state variables must be properly initialized. Observational metrics report the maximum sustained wind speed, which inherently contains asymmetric components induced by storm translation and environmental shear. Because the physical model predicts the symmetric inner-core wind speed, $v$, the observed maximum wind must be appropriately inverted. The net surface wind vector is modeled as the sum of the axisymmetric vortex wind, a latitudinally dependent fraction of the translation velocity, and a shear-induced isallobaric component \cite{emanuel2006statistical, emanuel2017}:
\begin{equation}
\mathbf{u}_{\text{net}} = \mathbf{V} + G\mathbf{u}_t + 0.1\mathbf{S}\frac{|\mathbf{V}|}{15},
\label{eq:unet}
\end{equation}
and
\begin{equation}
G = \min\left\{1, 0.8 + 0.35\left[1 + \tanh\left(\frac{\phi - 35}{10}\right)\right]\right\},
\label{eq:g_factor}
\end{equation}
where $\mathbf{u}_{\text{net}}$ is the net asymmetric surface wind vector, $\mathbf{V}$ is the axisymmetric azimuthal surface wind vector at the radius of maximum winds (whose magnitude is $v$), $\mathbf{u}_t$ is the storm translation velocity vector, $\mathbf{S}$ is the vertical wind shear vector, $G$ is an empirical weighting factor for the environmental background flow, and $\phi$ is the latitude in degrees. By setting the magnitude of the net wind equal to the observed maximum sustained wind ($|\mathbf{u}_{\text{net}}| = V_{\max}$), we can invert this relationship using a numerical bisection solver to obtain the required initial axisymmetric wind speed, $v$.
Following the wind initialization, the unobserved inner-core relative humidity, $m$, must also be specified. Rather than treating $m$ as a free parameter or estimating it via unconstrained neural networks, we initialize it through a physics-driven two-stage scheme. At 48 hours prior to the forecast start time ($t_0 = t_{\text{start}} - 48\text{ h}$), the initial moisture $m_0$ is analytically inverted directly from the intensity tendency.
Starting from $m_0$, the moisture evolution equation is integrated forward for 48 hours using observed intensity evolution and preliminary environmental forcing. This nudged spin-up period allows the moisture variable to relax into a physically balanced equilibrium state, $m(t_{\text{start}})$, ensuring thermodynamic consistency at the forecast launch time.

In addition to predicting the temporal evolution of TC intensity, the ensemble
forecast framework is used to construct spatially resolved probabilistic hazard
diagnostics. We consider two complementary quantities: the probability that the
TC center passes within a prescribed distance of a location and the probability
that the local surface wind speed exceeds a specified threshold during the
forecast period.

For ensemble realization $i$, let $\mathbf{r}_i(t)$ denote the predicted TC-center
position at forecast time $t$. The spatial track-strike probability at location
$\mathbf{x}$ is defined as

\begin{equation}
P_{\mathrm{strike}}(\mathbf{x})
=
\frac{1}{N}
\sum_{i=1}^{N}
\mathbb{I}
\left[
\min_{t\in\mathcal{T}}
d\left(\mathbf{x},\mathbf{r}_i(t)\right)
\leq R_s
\right],
\label{eq:strike_probability}
\end{equation}

where $N$ is the total number of ensemble realizations,
$d(\cdot,\cdot)$ denotes the great-circle distance,
$\mathcal{T}$ represents the forecast period, and
$R_s=75$~km is the prescribed strike radius.
Accordingly, $P_{\mathrm{strike}}(\mathbf{x})$ represents the fraction of
ensemble realizations for which the TC center passes within 75~km of
location $\mathbf{x}$ at any time during the forecast period.

To characterize the spatial distribution of wind hazards, each ensemble
realization is further mapped to a time-dependent local surface wind field,
$W_i(\mathbf{x},t)$, following the FHLO \cite{LinFHLO} wind-field formulation. The maximum
wind speed experienced at location $\mathbf{x}$ during the forecast period is

\begin{equation}
W_i^{\max}(\mathbf{x})
=
\max_{t\in\mathcal{T}}
W_i(\mathbf{x},t).
\label{eq:max_local_wind}
\end{equation}

For a prescribed wind-speed threshold $V_0$, the corresponding spatial
wind-speed exceedance probability is then defined as

\begin{equation}
P_{\mathrm{wind}}(\mathbf{x};V_0)
=
\frac{1}{N}
\sum_{i=1}^{N}
\mathbb{I}
\left[
W_i^{\max}(\mathbf{x}) \geq V_0
\right].
\label{eq:wind_exceedance_probability}
\end{equation}

Thus, $P_{\mathrm{wind}}(\mathbf{x};V_0)$ gives the ensemble probability
that the maximum local surface wind speed at location $\mathbf{x}$ exceeds
$V_0$ at least once during the forecast period. Unlike the track-strike
probability, which depends only on the storm-center trajectories, the
wind-speed exceedance probability propagates uncertainty in both storm
trajectory and intensity through the spatial wind-field calculation.
In this study, exceedance probabilities are evaluated for thresholds of
34, 50, 64, 83, and 96~kt.

\subsection{Environmental Data and Processing}
\label{Environmental Data and Processing}

Environmental predictors are obtained from ECMWF ERA5 reanalysis on a $0.25^\circ \times 0.25^\circ$ latitude--longitude grid at hourly temporal resolution \cite{hersbach2020era5}. Atmospheric predictor variables comprise horizontal wind components ($u$ and $v$), geopotential height ($z$), temperature ($T$), and specific humidity ($q$) across seven pressure levels (1000, 850, 700, 600, 500, 250, and 200~hPa). Surface boundary conditions include mean sea-level pressure ($MSLP$) and sea-surface temperature ($SST$), with the latter sampled 24 hours prior to the storm's arrival to ensure it represents the unperturbed pre-storm ocean state. Complementing the environmental data, storm tracks and intensity observations are sourced from the International Best Track Archive for Climate Stewardship (IBTrACS) for the North Atlantic and Eastern North Pacific basins, providing a rigorous basis for generalizability testing. The original 6-hourly best-track records are interpolated to an hourly resolution using cubic splines for spatial positions and shape-preserving piecewise cubic Hermite interpolating polynomials (PCHIP) for maximum sustained wind speed \cite{fritsch1980monotone}.
While interpolating 6-hourly best-track data provides the necessary temporal resolution for the dynamical solver, the unobserved inner-core moisture variable ($m$) can be highly sensitive to the exact intensity prescribed at the initial time step. To mitigate this initialization shock and the inherent coarseness of the observations, our framework employs a 48-hour dynamical spin-up period prior to the formal forecast integration. Although the absolute initial state at $t-48$ may contain interpolation errors, continuously forcing the system along the observed trajectory over a 48-hour window allows $m$ to adjust toward a dynamically balanced state with the intensity field and environmental thermodynamic forcing. Consequently, the state variables converge to a physically consistent manifold by the start of the prediction window ($t=0$), reducing the influence of initial transients and providing a robust initial condition for forecasting.

For each storm position, environmental fields are extracted onto a storm-centered $72 \times 72$ grid spanning $18^\circ \times 18^\circ$, or approximately a $\pm 1000$~km radius. This storm-relative representation removes geographic dependence and forces the neural closure to learn environmental controls that are physically linked to storm evolution rather than basin-specific spatial patterns, ultimately yielding a comprehensive three-dimensional representation of the thermodynamic and dynamical environment surrounding the cyclone.

In addition to historical single-track simulation experiments, FAST-ML is
comprehensively evaluated across three distinct probabilistic configurations
to systematically decouple track, environmental forcing, and intensity
closure errors.

First, for diagnostic evaluation under observed environmental conditions,
FAST-ML is driven by ERA5 reanalysis fields combined with ensemble cyclone
tracks obtained from the THORPEX Interactive Grand Global Ensemble (TIGGE)
archive for the ECMWF Ensemble Prediction System (EPS), which provides 51
member tracks at a 6-hourly resolution
\cite{bougeault2010tigge,swinbank2016tigge}. Second, to isolate intensity
closure performance from track prediction errors, the model is forced by the
same ERA5 environmental baseline combined with Google FNv3
\cite{alet2025skillful} forecast tracks. To adequately sample the spatial
uncertainty distribution and maintain methodological consistency across these
diagnostic evaluations, a per-lead-time conditional Gaussian Markov model is
fitted to the respective track velocities (ECMWF EPS and Google FNv3) to
generate expanded 100-member synthetic-track ensembles \cite{LinFHLO},
rather than relying solely on the raw forecast tracks. Third, to evaluate real-world operational forecasting capability, FAST-ML is deployed in a strict operational configuration. In this setup, the model is driven entirely by forecast fields and ensemble cyclone tracks from the NCEP
Global Ensemble Forecast System (GEFS), which provides 31 forecast members at a 6-hourly resolution. Consistent with the diagnostic configurations, the conditional Gaussian Markov track-velocity model is applied to the GEFS member tracks to construct 1,000 synthetic track realizations
\cite{LinFHLO}, allowing FAST-ML to yield probabilistic operational intensity
forecasts.

\subsection{Overview of the FAST-ML Framework}

FAST-ML introduces machine learning at a single, deliberately narrow point in the modeling chain: the parameterization of environmental ventilation. At each time step, the FAST-ML ingests an instantaneous three-dimensional environmental state,

\begin{equation}
e_t = [u_t, v_t, z_t, T_t, q_t,SST_t,MSLP_t],
\end{equation}

where $u_t, v_t$, and $z_t$ denote the 3D kinematic fields, $T_t$ and $q_t$ are the thermodynamic variables, and $SST_t$ and $MSLP_t$ serve as the surface boundary conditions as we discussed in section \ref{Environmental Data and Processing}, all derived from the gridded ERA5 data as detailed previously. The model then diagnoses the corresponding ventilation-related closure parameters,

\begin{equation}
\theta_t = [S_t, \chi_t].
\end{equation}

\begin{figure}[htbp]
    \centering
    \includegraphics[width=0.85\textwidth]{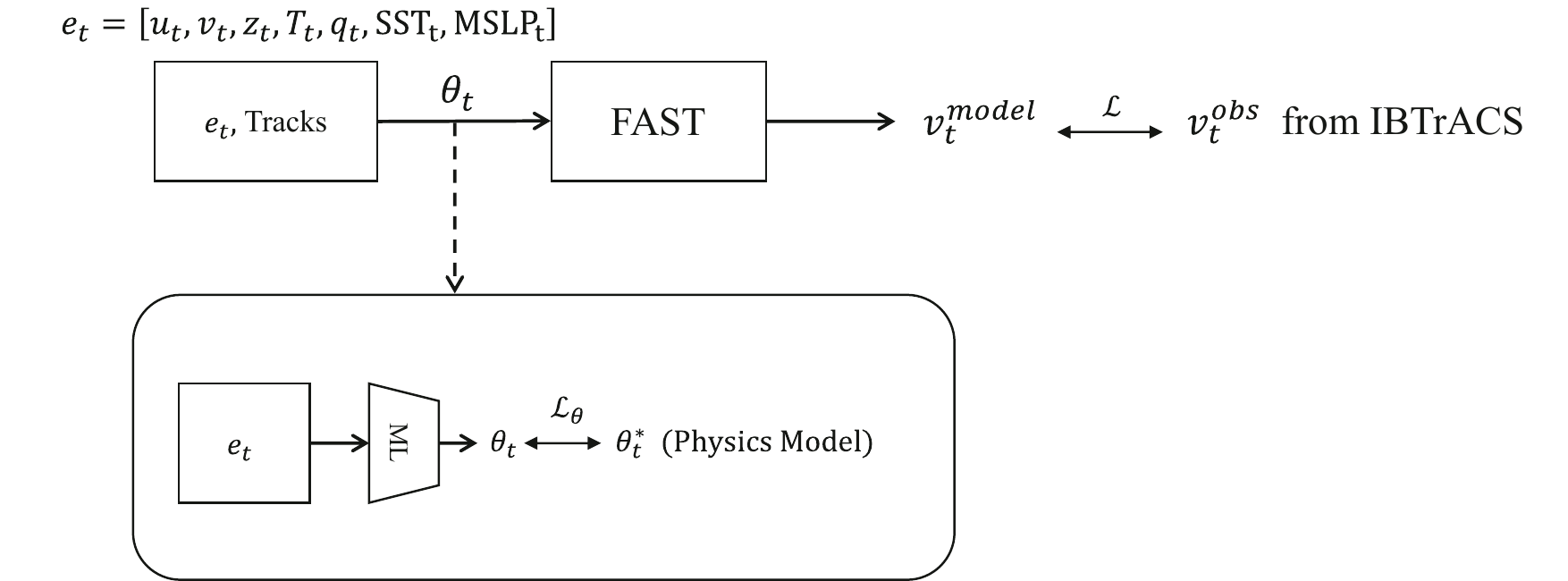} 
\caption{Overview of the FAST-ML hybrid physics--machine learning framework. Environmental fields ($e_t$) are processed by a neural closure model to estimate ventilation-related parameters ($\theta_t=[S_t,\chi_t]$), which are supplied to the FAST intensity model. The framework is trained through physics-guided pretraining and subsequent observation-driven end-to-end optimization using a differentiable FAST solver and IBTrACS intensity observations. By learning unresolved environmental forcing while retaining the governing thermodynamic equations, FAST-ML improves forecast skill while preserving physical consistency and interpretability. The downward dashed arrow illustrates the training process, where the parameter $\theta_t$ obtained from the physics model is passed down as the target label ($\theta_t^*$) to train the machine learning module.}    \label{fig:macro_logic}
\end{figure}

Rather than using these diagnosed quantities to predict intensity directly, $\theta_t$ is passed into the FAST governing equations, and storm evolution is obtained by numerically integrating Eqs.~(\ref{eq:dvdt})--(\ref{eq:dmdt}) forward in time. The neural network does not directly output intensity; it supplies environmental closure parameters and is trained through the physical solver using observed intensity trajectories.
This design establishes a strict division of labor between environmental inference and physical prediction. The FAST-ML is responsible solely for diagnosing the unresolved environmental forcing term $\chi S$, while the reduced-order physical model retains full responsibility for the temporal evolution of storm intensity. Every other thermodynamic quantity governing the cyclone heat engine, including the potential intensity $V_p$, the ocean-coupling factor $\alpha$, and the energetics coefficients $\beta$ and $\gamma$, retains its original analytical formulation and is left untouched by the learning process. Machine learning is therefore introduced only at the single location where the physical formulation is least constrained and most uncertain, namely the representation of environmental ventilation, rather than being used to substitute for physical relationships that are already well established.
The overall FAST-ML architecture is summarized in Figure~\ref{fig:macro_logic}. The framework couples a neural environmental closure to the FAST dynamical core through a differentiable optimization strategy: environmental fields are mapped to the ventilation-related closure parameters, which are then injected directly into the physical model's governing equations at every integration step. This coupling is trained in two stages, beginning with physics-guided pretraining and followed by observation-driven end-to-end optimization.

\subsection{Neural Environmental Closure Model}

\subsubsection{Three-Dimensional Environmental Representation}

\begin{figure*}[htbp]
    \centering

    \includegraphics[width=1.1\textwidth]{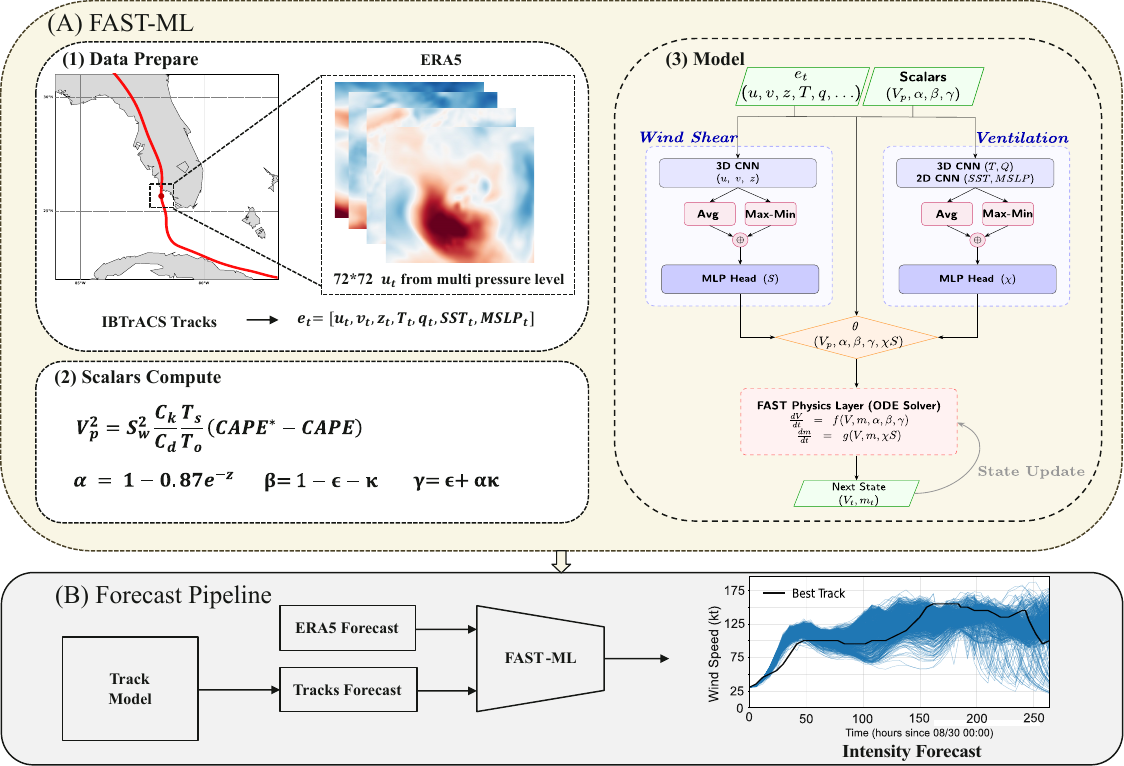} 
\caption{Architecture of the FAST-ML environmental closure and forecasting framework. (A) Multilevel environmental fields are processed by a physically informed dual-stream CNN. The kinematic stream estimates environmental shear ($S$), while the thermodynamic stream estimates mixing efficiency ($\chi$). Average (Avg) and Max--Min pooling capture large-scale conditions and localized asymmetries. FAST thermodynamic parameters ($V_p$, $\alpha$, $\beta$, and $\gamma$) are computed analytically. (B) Coupling of the learned environmental closure with the differentiable FAST solver for TC intensity forecasting. (Note: MLP is multilayer perceptron.)}
\label{fig:detailed_architecture}
\end{figure*}

The environmental closure model receives multilevel atmospheric fields describing the thermodynamic and dynamical structure surrounding the TC. Atmospheric variables are represented as three-dimensional tensors spanning seven pressure levels. This allows the model to exploit vertical structure and identify the atmospheric layers most relevant to storm-environment interaction.
Providing the complete vertical profile eliminates the need to prescribe a priori which pressure levels govern ventilation. Instead, the network is allowed to learn the vertical structures that best explain observed intensity evolution directly from data.
The detailed architecture of the FAST-ML is shown in Figure~\ref{fig:detailed_architecture}. The framework employs a dual-stream design that separately represents the dynamical and thermodynamic components of ventilation while simultaneously incorporating analytically derived thermodynamic scalars required by the FAST intensity model.

\subsubsection{Dual-Stream Architecture}

Ventilation arises from the interaction of two physically distinct mechanisms: dynamical forcing through vertical wind shear and thermodynamic susceptibility through environmental humidity and entropy structure. Because a single scalar ventilation parameter cannot distinguish which of these two mechanisms is responsible for a given forcing magnitude, estimating it as one lumped quantity therefore introduces an equifinality problem, whereby different combinations of shear and moisture deficit can produce the same net ventilation magnitude.
To address this challenge, FAST-ML employs a physically informed dual-stream architecture, in which separate network branches infer the dynamical and thermodynamic contributions to ventilation independently before they are recombined through the physical model.
The dynamical (kinematic) stream receives geopotential height and horizontal wind fields $(u,v,z)$ and predicts the effective shear magnitude, $S$. These variables encode information related to environmental steering flow, vertical shear, and synoptic-scale dynamical forcing.The thermodynamic stream, which predicts the mixing efficiency $\chi$, is structurally bifurcated to handle different input dimensionalities. A 3D sub-stream processes the three-dimensional atmospheric thermodynamic structure (temperature, $T$, and specific humidity, $q$), while a parallel 2D sub-stream independently encodes the surface boundary conditions (mean sea-level pressure, $MSLP$ and sea-surface temperature, $SST$). The latent representations from both sub-streams are subsequently concatenated to merge the atmospheric profile with surface forcing, yielding the final $\chi$ prediction. 

This separation allows the network to preserve the distinct physical roles of shear forcing, atmospheric moisture profiles, and surface fluxes, which already enter the FAST equations as independent multiplicative terms in the ventilation product $\chi S$, while reducing parameter degeneracy.

\subsubsection{Implicit Vortex Removal}
Calculating the true environmental vertical wind shear conventionally requires the explicit mathematical removal of the TC-induced circulation from the total wind field. A standard approach, used in the original FAST model, sets the relative vorticity ($\zeta$) and divergence ($d$) to zero within a rigidly defined inversion radius (e.g., $r^* = 400$~km) of the storm center. The environmental wind field ($\mathbf{u}_{env}$) is then isolated by inverting the Poisson equation on a sphere to obtain the vortex-associated streamfunction ($\psi_{vortex}$) and velocity potential ($\Phi_{vortex}$):
\begin{equation}
\nabla^2 \psi_{vortex} = \zeta_{vortex}, \quad \nabla^2 \Phi_{vortex} = d_{vortex},
\end{equation}
\begin{equation}
\mathbf{u} = \mathbf{u}_{env} + \nabla \Phi_{vortex} + \hat{k} \times \nabla \psi_{vortex}.
\end{equation}
While physically intuitive, this traditional vortex-removal technique introduces severe computational bottlenecks: solving the Laplacian operator ($\nabla^2$) globally, or even over expansive regional grids, at every time step is computationally expensive.
The traditional approach is also constrained by rigid empirical assumptions, including a static inversion radius ($r^*$), fixed pressure levels for shear calculation (e.g., 250~hPa and 850~hPa), and the idealized assumption that the background environment is irrotational and non-divergent. The latter is particularly problematic at upper levels (e.g., 250~hPa), where intense tropical cyclone outflow, upper-level troughs, and jet streams introduce strong environmental divergence and vorticity. These assumptions are poorly suited to the continuous structural evolution, varying sizes, and complex high-level synoptic interactions of real-world TCs. FAST-ML avoids both the computational overhead and these restrictive assumptions by leveraging the spatial feature-extraction capability of its 3D convolutional neural network (CNN). Rather than performing an explicit Poisson inversion, the network ingests the raw, unfiltered 3D kinematic fields ($u$, $v$, $z$) directly and learns representations of environmental shear from the full atmospheric structure. This allows the model to represent complex three-dimensional environmental features, including upper-level outflow channels and trough dynamics, without requiring potential-flow assumptions.

Notably, FAST-ML, operating on these unfiltered input fields, produces intensity forecasts that are more accurate than the original FAST model using explicitly filtered environmental winds. This result indicates that the network not only recovers an implicit vortex-separation capability, but does so in a way that outperforms the rigid mathematical filter it replaces. By learning a spatially adaptive, continuously evolving representation of environmental shear, FAST-ML achieves both a substantial reduction in computational runtime and an improvement in forecast accuracy.

\subsubsection{Feature Pooling and Closure Prediction}
Each stream consists of a series of three-dimensional convolutional layers that extract multiscale environmental features. Following feature extraction, both global average pooling and Max--Min pooling are applied: average pooling summarizes the large-scale synoptic environment, while Max--Min pooling characterizes localized spatial extremes and asymmetries.
This dual-pooling design is directly motivated by the analytical formulations of TC ventilation. Traditional physical schemes diagnose the environmental entropy deficit largely through extreme-value statistics rather than simple spatial means: to quantify detrimental midlevel downdrafts, conventional parameterizations typically evaluate the gridpoint maximum of saturation deficits and extract high upper-tail percentiles. Central-tendency quantities---such as the median air--sea thermodynamic disequilibrium used to scale the deficit---are instead naturally approximated by the average-pooling branch.
Conventional parameterizations, however, remain fundamentally limited by their reliance on rigid, empirical thresholds. Parameters such as the diagnostic percentile ($N$), calculation radius ($r_{env}$), and deficit cap ($x_d$) are assigned fixed, basin-specific constants (e.g., $N=90$ and $r_{env}=1000$~km for the Atlantic versus $N=50$ for the eastern Pacific). This static hardcoding limits the ability to adapt to anomalous synoptic environments, varying storm sizes, or rapidly evolving structural changes.
The proposed architecture overcomes these limitations. By combining Max--Min and average pooling, the network retains the physical philosophy of balancing extreme spatial anomalies against mean background states, without being constrained by fixed percentiles or rigid spatial boundaries. The network instead learns to dynamically isolate the most critical thermodynamic extremes from instantaneous, high-resolution environmental data. The resulting feature vectors map to final estimates of $S$ and $\chi$ that remain physically grounded, while being more precise, adaptive, and dynamically responsive than traditional parameterizations.

\subsection{Differentiable Physical Integration and Training}

The FAST-ML is coupled to the FAST solver through a differentiable ordinary differential equation (ODE) framework \cite{chen2018neural}. To enable end-to-end gradient propagation from trajectory prediction errors back to the FAST-ML parameters, the governing differential equations are integrated using a differentiable Heun's second-order predictor-corrector scheme (explicit trapezoidal rule). For each 1-hour model step, the diagnosed closure parameters $(\hat{S}_t, \hat{\chi}_t)$ are held piecewise constant while the physical state is advanced using an internal numerical sub-step of $\Delta t = 0.25$~h. Consequently, gradients propagate through every integration step of the physical model via Backpropagation Through Time (BPTT), allowing closure parameters to be optimized directly against observed storm evolution.
The inner-core moisture variable, $m$, is treated as a latent physical state. Because reliable observations of $m$ are unavailable at the temporal frequency required for training, it is not directly supervised. Instead, its trajectory evolves according to Eq.~(\ref{eq:dmdt}), and learning occurs indirectly through its effect on storm intensity.

\subsubsection{Physics-Guided Pretraining}

Optimization is performed in two stages using the Adam optimizer, coupled with a cosine annealing learning rate schedule and a gradient clipping norm of 1.0 to ensure stable convergence. During the first stage, the FAST-ML is initialized using pseudo-labels derived from the original FAST parameterizations. Specifically, conventional shear diagnostics and the entropy-deficit estimate $\chi_{\rm grid}$ are used as supervision targets. 
The purpose of this stage is not to reproduce the analytical closure itself. Rather, it constrains the network within a physically consistent manifold and prevents the differentiable solver from exploring unstable regions of parameter space during early optimization.

\subsubsection{Observation-Driven End-to-End Optimization}

After pretraining, pseudo-label supervision is removed and the entire framework is optimized directly against observed intensity evolution. The loss function is defined as

\begin{equation}
\mathcal{L}
=
\sum_{t=0}^{T}
{\rm Huber}
\left(
v_t^{\rm model},
v_t^{\rm obs}
\right)
+
\lambda_\theta
\sum_{t=0}^{T-1}
(\theta_{t+1}-\theta_t)^2,
\end{equation}

where $v_t^{\rm model}$ and $v_t^{\rm obs}$ denote simulated and observed intensity, respectively. The second term imposes temporal smoothness on the closure parameters and suppresses unrealistic high-frequency fluctuations in the diagnosed environmental forcing. 
Through this two-stage training strategy, the FAST-ML is first anchored to physically meaningful solutions and subsequently allowed to depart from imperfect analytical assumptions when required to better reproduce observed storm evolution.

\subsection{Model Interpretability}

To investigate the physical mechanisms learned by FAST-ML, we perform gradient-based attribution analysis on the trained FAST-ML. Saliency maps are computed with respect to both closure parameters, $S$ and $\chi$, by evaluating the gradient of each parameter with respect to the environmental input variables.
For each storm, gradients are aggregated spatially, vertically, and temporally to quantify the relative importance of individual variables and pressure levels. 
Furthermore, to rigorously validate the physical realism of the learned parameterization, we evaluate the physical correlation between the net ventilation variable predicted by the model---derived explicitly as the multiplicative product of mixing efficiency and effective wind shear magnitude ($\chi \cdot S$)---and the actual environmental ventilation, calculated as the surface integral of eddy entropy flux across the inner-core boundary ($r \approx 200\text{~km}$) as we introduced in section \ref{FAST}.
Together, these analyses provide a physically interpretable diagnosis of which environmental structures most strongly influence the learned ventilation parameterization, verifying that FAST-ML captures true physical mechanisms in direct alignment with established theories of TC ventilation and shear-induced suppression.

\subsection{Evaluation Methodology}
\label{evaluate}

Deterministic forecast performance is evaluated using Root Mean Square Error (RMSE) and Mean Absolute Error (MAE) relative to IBTrACS observations.
Probabilistic forecast performance is evaluated using the Continuous Ranked Probability Score (CRPS), which measures the agreement between forecast distributions and observed outcomes.
RI events are defined following \cite{kaplan2010} as a 24-hour increase in maximum sustained wind speed of at least 30 kt ($\Delta V_{24} \geq 30$ kt). Forecast skill for RI events is evaluated using confusion-matrix statistics, including the Probability of Detection (POD), False Alarm Ratio (FAR), Critical Success Index (CSI), Peirce Skill Score (PSS), and Frequency Bias.
The framework is strictly trained on North Atlantic TC data from 2003--2021, with the 2022--2023 seasons held out as a validation set used to supervise model training and tuning. The independent 2024 season is reserved as a fully out-of-sample test period and is not used at any stage of model development. Furthermore, to evaluate the model's operational relevance and compare it against the latest state-of-the-art AI weather models, selected high-impact storms from the most recent 2025 season are introduced as independent, extended real-time case studies. Storms from 2025 are used only for additional out-of-sample evaluations and case studies, and are not included in model training or hyperparameter tuning.

To establish a state-of-the-art probabilistic baseline for the aforementioned comparative configurations, we incorporate Google DeepMind's FNV3 \cite{alet2025skillful}, a global AI weather system that generates joint ensemble forecasts from deterministic marginals. FNV3 is trained extensively on decades of high-resolution ERA5 global reanalysis data to simulate comprehensive atmospheric evolution.

The ensemble size is experiment-specific. For the diagnostic
probabilistic experiments and comparison with FNV3, FAST-ML uses
100-member ensembles. For the GEFS-based operational forecasting
experiments, 1,000 synthetic track realizations are generated to
construct the corresponding probabilistic intensity forecasts.

FAST-ML is evaluated over 23 years (2003--2025) of TC activity through five complementary analyses: (1) multidecadal deterministic forecast performance across the training, validation, and independent test periods; (2) storm-scale examination of lifecycle evolution and physical consistency; (3) probabilistic forecasting skill benchmarked against Google DeepMind's FNV3, including cross-basin zero-shot evaluation in the Eastern Pacific; (4) architectural ablation experiments designed to isolate the contribution of individual model components; and (5) computational efficiency and practical scalability relative to modern AI forecasting systems. Together, these analyses are designed to assess not only intensity forecast accuracy, but also physical realism, robustness, generalizability, and applicability to large-ensemble TC intensity prediction.

\section{Results}
Following the evaluation design outlined in Section~\ref{evaluate}, we first assess long-term statistical performance across the training, validation, and independent test periods.

\subsection{Long-Term Statistical Performance and Robustness}
\label{single track}
\begin{figure}[H]
    \centering
    \includegraphics[width=7in]{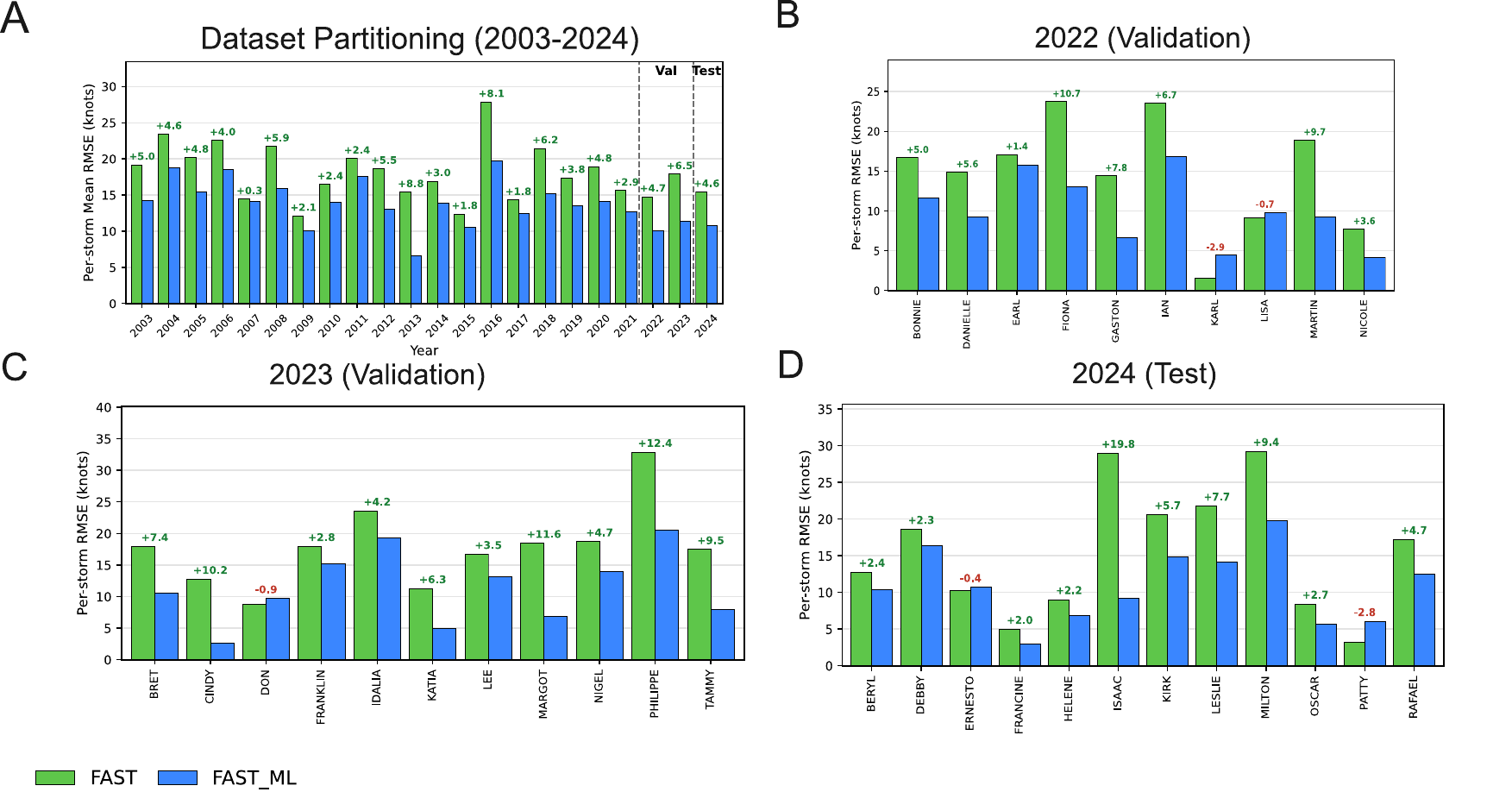}
\caption{Deterministic intensity forecast performance and dataset partitioning for FAST and FAST-ML. (A) Annual mean per-storm RMSE (knots) from 2003 to 2024. Dashed vertical lines denote the chronological dataset split into training (2003--2021), validation (2022--2023), and independent test (2024) periods. (B--D) Per-storm RMSE for individual named storms during the 2022 (B) and 2023 (C) validation seasons, and the 2024 test season (D). Green and blue bars represent baseline FAST and FAST-ML forecast errors, respectively. Numerical annotations indicate the net RMSE reduction (FAST minus FAST-ML), where positive values denote performance improvement by the FAST-ML.}
    \label{fig:macro_stats}
\end{figure}

Figure~\ref{fig:macro_stats} summarizes per-storm mean RMSE by year across the North Atlantic basin. A central requirement for any machine-learning-enhanced physical model is its ability to remain skillful across a wide range of environmental conditions while generalizing beyond the storms used for training. Across the 2003--2021 training period and the 2022--2023 validation seasons, FAST-ML reduces per-storm RMSE relative to the baseline FAST model in nearly every year (Fig.~\ref{fig:macro_stats}A). The largest improvements occur during seasons containing storms for which the fixed analytical ventilation closure produces its largest forecast errors, indicating that the learned environmental parameterization is particularly effective at correcting situations where the static closure becomes inadequate (Fig.~\ref{fig:macro_stats}B--C). In the 2023 validation season, for example, the FAST-ML reduces RMSE by 12.4 kt for Hurricane Philippe (2023), characterized by baseline FAST errors exceeding 30--50 kt.

This improvement persists in the independent 2024 test season, although with a smaller basin-mean magnitude (Fig.~\ref{fig:macro_stats}D). Across the 12 storms in this held-out set, FAST-ML reduces the per-storm mean RMSE by 4.6 kt relative to FAST. The largest individual improvements occur for Hurricane Isaac (2024), for which RMSE decreases by 19.8 kt. For Helene and Milton, two of the most damaging storms of the 2024 season, the improvements are more modest but remain positive, with RMSE reductions of 2.2 kt and 9.4 kt. These results indicate that the primary benefit of the FAST-ML is not a uniform reduction of error across all storms, but rather a targeted reduction of the largest forecast failures associated with misrepresentation of environmental ventilation. Improvements vary among storms, and FAST-ML slightly increases RMSE for Ernesto and Patty.

\begin{figure}[H]
    \centering
    \includegraphics[width=\textwidth]{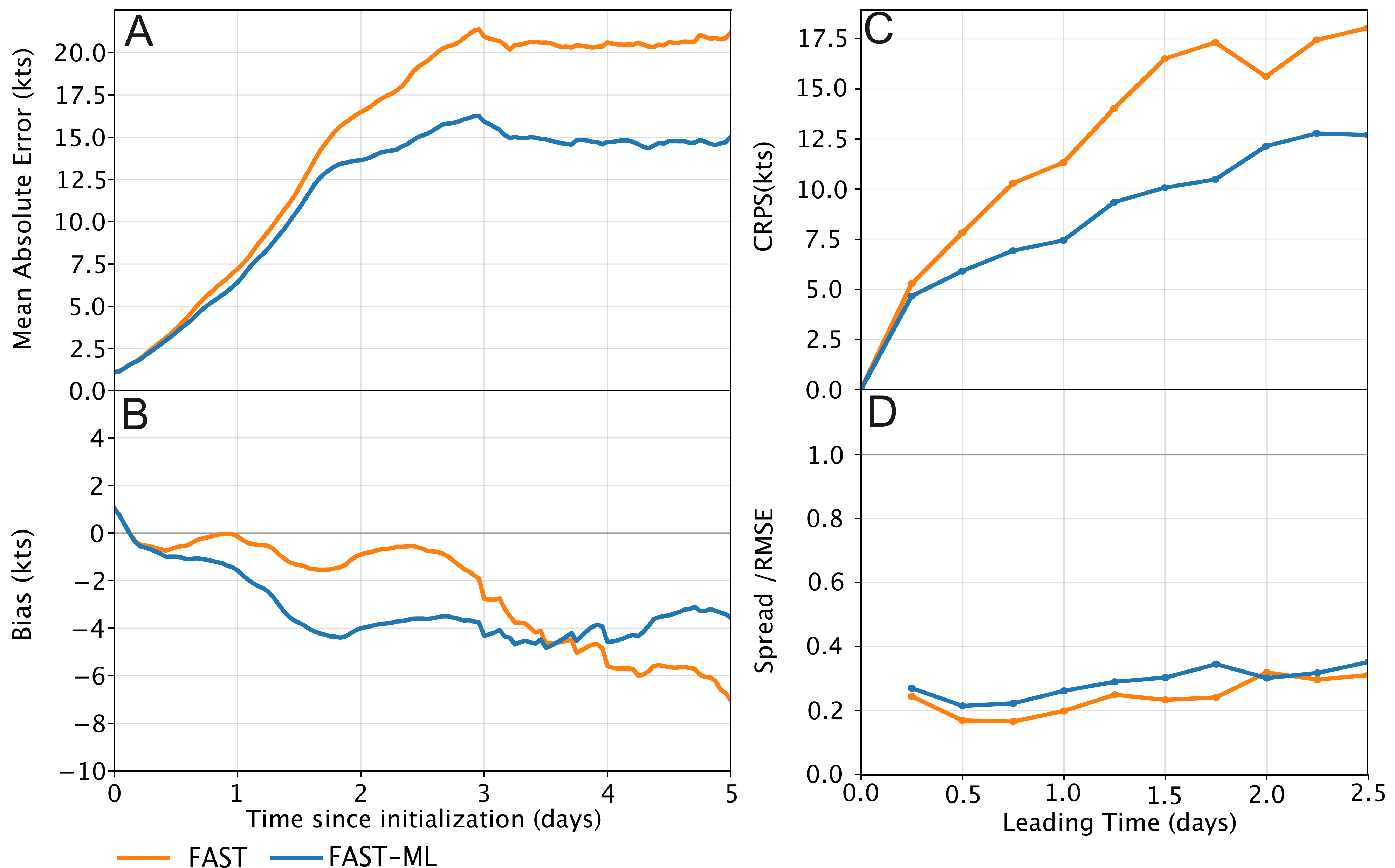}
\caption{Deterministic and probabilistic forecast performance as a function of lead time.
(A) MAE evaluated along observed best-track trajectories for North Atlantic storms over a 5-day horizon.
(B) Systematic intensity bias.
(C) CRPS for ensemble forecasts evaluated over a 2.5-day horizon, illustrating the improved probabilistic skill of FAST-ML. (D) Ensemble spread-to-RMSE ratio. Test set includes forecasts from 2024 to 2025 in the North Atlantic basin and only forecasts where the initial intensity is greater than $30ms^{-1}$ are included in
the samples. Most of ECMWF tracks landed after 2.5 days so we choose the first 2.5 days for ensemble evaluation.}
    \label{fig:mae_bias_comparison}
\end{figure}

Annual RMSE statistics provide a storm-integrated view of performance but do not reveal how forecast errors accumulate through time. To examine the temporal evolution of forecast skill, we evaluate both deterministic and probabilistic errors as a function of lead time (Figure~\ref{fig:mae_bias_comparison}). For deterministic forecasts evaluated along observed best-track trajectories, both FAST and FAST-ML begin from the same initialization error, but their subsequent error growth differs substantially. The MAE of the baseline FAST model increases approximately monotonically with forecast lead time, exceeding 20 kt by day 5. In contrast, the MAE of FAST-ML grows more slowly and levels off near 15 kt after approximately day 3 (Figure~\ref{fig:mae_bias_comparison}A). The bias evolution provides additional insight: FAST develops a persistent negative bias, reaching approximately -6 to -8 kt by day 5, consistent with excessive environmental suppression. FAST-ML also retains a weak negative bias, but its magnitude stabilizes closer to -3 to -4 kt over the same forecast window (Figure~\ref{fig:mae_bias_comparison}B). This suggests that the improved closure formulation moderates the cumulative effects of environmental ventilation errors during storm evolution.

Beyond deterministic skill, we extend our evaluation to the probabilistic performance of the framework using ensemble forecasts over a 2.5-day horizon. As shown in Figure~\ref{fig:mae_bias_comparison}C, FAST-ML yields a substantially lower CRPS compared to the baseline, maintaining an error of approximately 12.5 kt at 60 hours versus nearly 18 kt for FAST. This indicates a marked improvement in overall probabilistic forecast skill. However, evaluating the ensemble calibration via the Spread-to-RMSE ratio (Figure~\ref{fig:mae_bias_comparison}D) reveals that both systems are significantly under-dispersive, with ratios remaining between 0.2 and 0.35. While FAST-ML exhibits a marginally higher (improved) spread ratio than FAST, the persistent under-dispersion indicates that the input perturbations do not fully overcome the strong deterministic constraints of the underlying physical ODEs. Despite this under-dispersion, the FAST-ML significantly tightens the forecast distribution closer to the true observation, as evidenced by the concurrent reduction in both MAE and CRPS. Future work will explore additional perturbation strategies and statistical calibration to improve ensemble reliability.

This same behavior is reflected in RI classification skill (Table~\ref{tab:ri_performance}). Relative to baseline FAST, FAST-ML substantially reduces the FAR from 0.400 to 0.220 and increases the CSI from 0.304 to 0.351. Crucially, this massive reduction in false alarms (False Positives dropped from 30 to 13) is achieved without sacrificing sensitivity; in fact, the POD slightly improves from 0.381 to 0.390 (True Positives increased from 45 to 46). While the baseline FAST model exhibits a Frequency Bias (0.636) closer to the ideal value of 1.0 compared to FAST-ML (0.500), this higher bias in the baseline is artificially inflated by its large number of spurious false alarms. The lower bias in FAST-ML does not represent a degradation in forecasting actual RI events, but rather demonstrates that the model has become significantly more precise and selective. This result is consistent with the interpretation that FAST-ML improves forecast skill by better regulating the environmental conditions under which RI can be sustained, effectively filtering out structurally unfavorable environments that the baseline analytical closure incorrectly flagged as RI.

\begin{table}[H]
\centering
\caption{RI ($\Delta V_{24} \geq 30$~kts) classification performance. The evaluation combines the Confusion Matrix (TP: True Positive, FP: False Positive, FN: False Negative, TN: True Negative) and Skill Metrics (POD: Probability of Detection, FAR: False Alarm Ratio, CSI: Critical Success Index, PSS: Peirce Skill Score, Bias: Frequency Bias).}
\label{tab:ri_performance}
\begin{tabular}{l cccc c ccccc}
\toprule
& \multicolumn{4}{c}{Confusion Matrix} && \multicolumn{5}{c}{Skill Metrics} \\
\cmidrule{2-5} \cmidrule{7-11}
& TP & FP & FN & TN && POD & FAR & CSI & PSS & Bias \\
\midrule
Baseline FAST  & 45 & 30 & 73 & 1364 && 0.381 & 0.400 & 0.304 & 0.360 & \textbf{0.636} \\
FAST-ML        & \textbf{46} & \textbf{13} & \textbf{72} & \textbf{1381} && \textbf{0.390} & \textbf{0.220} & \textbf{0.351} & \textbf{0.381} & 0.500 \\
\bottomrule
\end{tabular}
\end{table}

\subsection{Model Capabilities in Capturing Complex Lifecycle Dynamics}

\begin{figure}[H]
    \centering
    \includegraphics[width=0.95\textwidth]{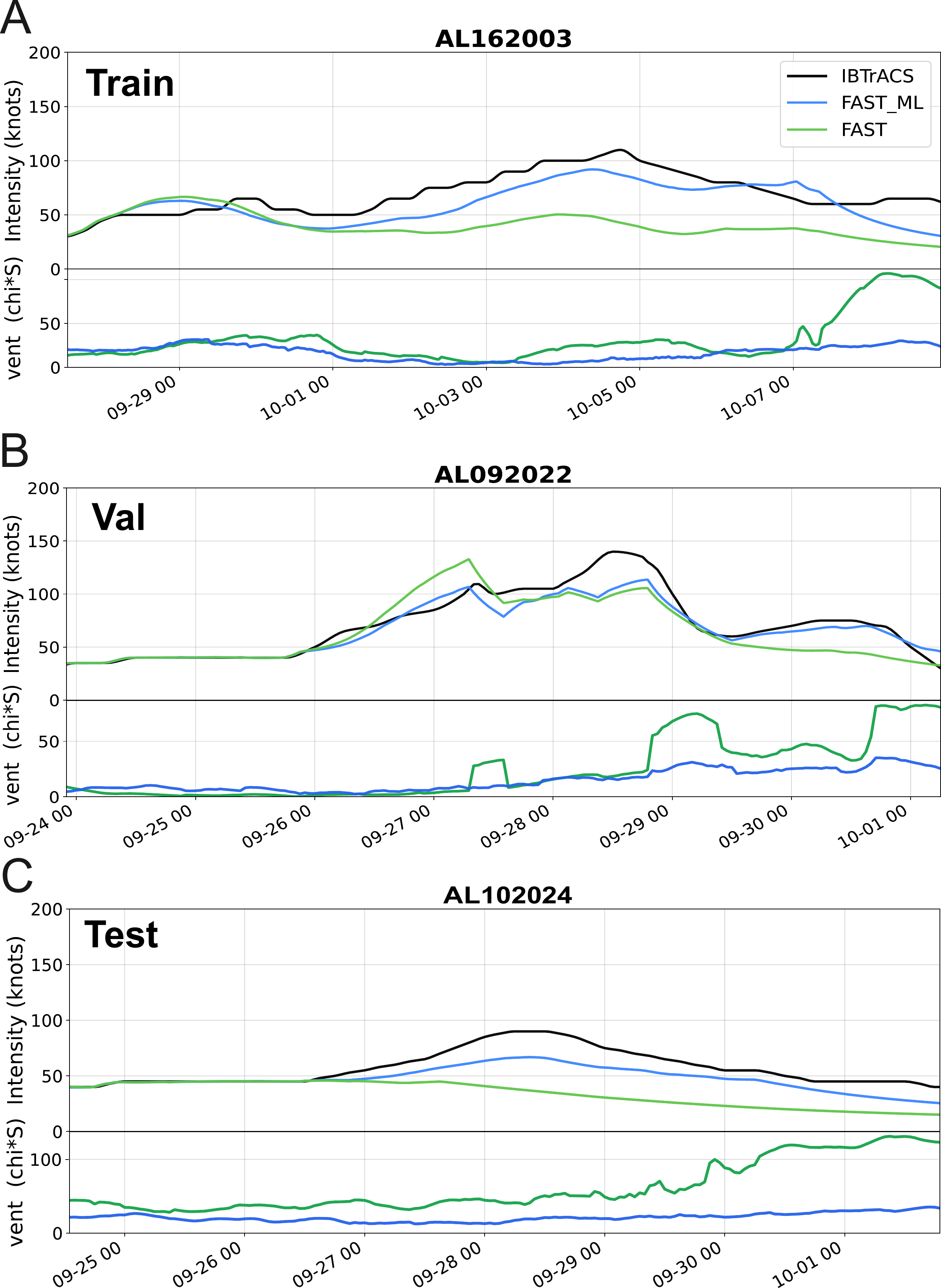}
    \caption{Lifecycle tracking and dynamic ventilation regulation by the FAST-ML framework. The panels illustrate the temporal evolution of maximum sustained wind speed (top subpanels) and the effective ventilation parameter ($\chi S$, bottom subpanels) for three representative TCs across the data splits: (A) Hurricane Kate (2003) from the Train set, (B) Hurricane Ian (2022) from the Validation set, and (C) Hurricane Isaac (2024) from the Test set. The baseline FAST model (green) frequently exhibits structural errors—such as spurious over-intensification or premature dissipation—due to rigid analytical ventilation closures. In contrast, FAST-ML (blue) dynamically adjusts the environmental forcing, maintaining physically consistent parameter regimes that faithfully track the observed IBTrACS intensity (black) across diverse and complex lifecycle phases.}
    \label{fig:lifecycle_dynamics}
\end{figure}

While aggregate statistics quantify forecast skill at the basin scale, they do not reveal the underlying mechanisms responsible for model improvement. To better understand how the FAST-ML modifies storm evolution, we examine representative lifecycle simulations spanning distinct intensity regimes and forecast failure modes (Figure~\ref{fig:lifecycle_dynamics}).
A key limitation of the original FAST formulation is that the analytical ventilation closure can occasionally force the moisture equation into unrealistic regimes, either suppressing storm development too aggressively or permitting excessive intensification. Because ventilation directly regulates the evolution of inner-core moisture, even moderate errors in the closure can accumulate over time and produce substantial intensity biases. The case studies presented in Figure~\ref{fig:lifecycle_dynamics} illustrate how FAST-ML mitigates both behaviors through a dynamically evolving environmental parameterization.

The first failure mode is chronic under-intensification and premature storm weakening caused by excessive ventilation, illustrated by the training set example Hurricane Kate (2003) (Fig.~\ref{fig:lifecycle_dynamics}A). While this storm was included in the model's training distribution, analyzing its lifecycle provides a crucial mechanistic baseline to understand how the neural closure structurally alters the environmental forcing before assessing out-of-sample generalization. In the baseline FAST simulation, the analytical ventilation term remains elevated and eventually spikes to extreme, unphysical values late in the lifecycle. This persistent overestimation of environmental suppression drives the inner-core moisture state toward unrealistically low values, limiting intensification and causing the simulated storm to remain substantially weaker than the observed IBTrACS peak. In contrast, FAST-ML diagnoses a considerably smaller and more stable ventilation magnitude throughout this period, maintaining sufficient inner-core moisture to support continued development. The resulting intensity evolution follows the observed trajectory more closely, demonstrating the fundamental capability of the hybrid framework to accurately constrain environmental ventilation when exposed to known storm dynamics.

A second failure mode arises when the analytical closure underestimates environmental regulation, allowing unrealistically strong intensification followed by erratic corrections. Hurricane Ian (2022) from the validation set (Fig.~\ref{fig:lifecycle_dynamics}B) provides an example of this behavior. In the baseline FAST simulation, the ventilation term remains too weak prior to the observed peak, producing a severe early intensity overshoot relative to observations. This over-intensification is subsequently met with abrupt, erratic spikes in the diagnosed ventilation term, triggering an unphysical collapse in storm intensity. FAST-ML predicts a ventilation evolution with a much more stable and realistic magnitude, thereby maintaining the thermodynamic constraint imposed by environmental forcing and preventing excessive vortex amplification while smoothly tracking the double-peak structure.

A related manifestation of extreme environmental suppression is evident in the independent test set for Hurricane Isaac (2024) (Fig.~\ref{fig:lifecycle_dynamics}C). Here, the baseline FAST model exhibits a massive, runaway increase in the diagnosed ventilation term, which artificially caps the storm's intensity and triggers premature storm weakening while the observed system continues to strengthen. Such behavior suggests that the analytical closure becomes overly sensitive and produces runaway values when exposed to certain complex environmental states. FAST-ML instead produces a significantly lower, smoother ventilation trajectory, yielding a much closer correspondence with the observed intensity evolution throughout the storm lifecycle and preventing spurious dissipation.

Taken together, these case studies indicate that the primary contribution of the FAST-ML is not simply a systematic increase or decrease in ventilation, but rather a dynamic regulation that keeps the environmental forcing within a physically plausible range across diverse storm environments. This regulation reduces both over-suppression and under-suppression of storm growth, thereby improving intensity forecasts through a more balanced representation of storm--environment interaction. Importantly, the examples shown here are consistent with the basin-scale bias reductions reported in Section~3.1, suggesting that improved forecast skill emerges from correcting a recurring structural deficiency in the ventilation parameterization rather than from isolated case-specific improvements.

\subsection{Probabilistic Forecasting and Benchmark Comparisons against Google FNV3}
\label{ensemble forecast}

Recent advancements in global AI weather modeling have demonstrated that highly skillful ensemble forecasts can be generated efficiently by modeling joint distributions from deterministic marginals. Traditional ensemble generation often requires running computationally expensive, high-resolution global models multiple times. In contrast, recent methodologies such as Google DeepMind’s FNV3 circumvent this bottleneck by leveraging advanced statistical or generative techniques to capture the complex spatial, temporal, and cross-variable dependencies inherent in atmospheric dynamics. By transforming independent marginal predictions into physically coherent joint ensembles, this framework achieves state-of-the-art probabilistic skill at a fraction of the computational cost. FNV3 represents one of the most advanced AI-based probabilistic weather forecasting systems currently available and serves as a relevant benchmark for evaluating probabilistic forecast skill \cite{Dem24, alet2025skillful}.
While the deterministic experiments demonstrate that the FAST-ML improves intensity prediction along observed storm trajectories, operational forecasting requires accurate characterization of uncertainty in addition to accurate prediction of the most likely intensity evolution. We therefore evaluate FAST-ML in an ensemble forecasting configuration and compare its performance against FNV3. This comparison should be interpreted as an assessment of complementary forecasting approaches rather than a direct equivalence between model classes, as FNV3 is a global AI forecasting system that predicts atmospheric evolution, whereas FAST-ML is a specialized, physics-constrained model for along-track TC intensity prediction. Within this narrower intensity-forecasting task, the comparison provides a useful benchmark for evaluating whether a computationally efficient hybrid framework can achieve competitive probabilistic skill while retaining physical interpretability.

To systematically evaluate the framework, we assess FAST-ML across three progressively challenging configurations: (1) a baseline diagnostic setup using ERA5 fields, (2) a hybrid setup driven by external AI-generated track samples (Google DeepMind’s FNV3), and (3) a strict real-time operational setup driven solely by GEFS forecast fields and tracks.

\begin{figure}[H]
    \centering
    \includegraphics[width=0.95\textwidth]{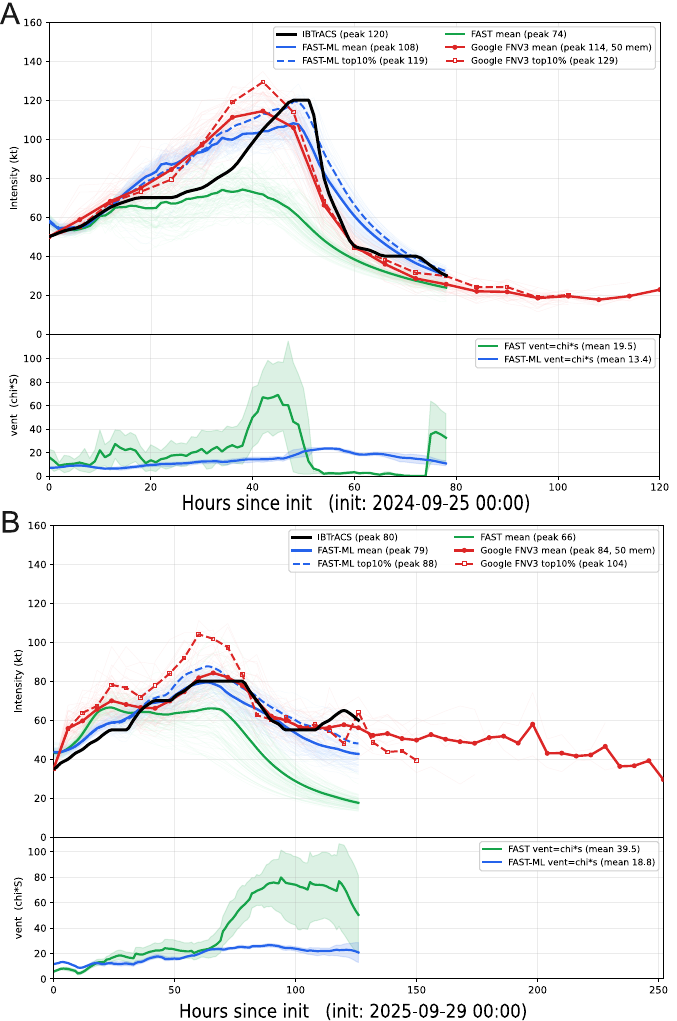}
    \caption{Probabilistic ensemble performance of FAST-ML versus Google FNV3 in North Atlantic. Ensemble intensity forecasts for (A) Hurricane Helene (2024) and (B) Hurricane Imelda (2025). Across both cases, FAST-ML (blue) closely tracks the forecast skill of the Google FNV3 benchmark (red) and IBTrACS observations (black), capturing both major and moderate Atlantic intensification regimes.}
    \label{fig:atlantic_ensemble_baseline}
\end{figure}

While the 2024 test set establishes the baseline physical consistency and statistical skill of FAST-ML, benchmarking against state-of-the-art global AI forecasting systems requires evaluation on recent, complex, and high-impact events. Therefore, we extend our analysis to include extreme tropical cyclones from the 2024 season as well as out-of-sample events from the 2025 season.
Hurricane Helene (2024)(Fig.~\ref{fig:atlantic_ensemble_baseline}A) provides a more challenging test case. FAST-ML underpredicts the observed peak intensity of 120 kt, with an ensemble-mean maximum of 108 kt, whereas the FNV3 ensemble mean reaches 114 kt and is therefore closer to the observed intensity. However, analysis of the ensemble distribution reveals a different perspective. The upper-decile FAST-ML members reach approximately 119 kt, nearly matching the observed peak, whereas the upper tail of FNV3 extends to roughly 129 kt. This suggests that, although the central tendency of FAST-ML exhibits a larger low bias, its ensemble distribution remains closely aligned with the observed outcome. These results highlight the importance of evaluating both ensemble means and forecast distributions when assessing probabilistic forecasting systems.
For Hurricane Imelda (2025)
(Fig.~\ref{fig:atlantic_ensemble_baseline}B), FAST-ML and FNV3 exhibit comparable forecast skill. The FAST-ML ensemble mean reaches 79 kt, closely matching the observed peak intensity of 80 kt and demonstrating skill similar to that of FNV3. An additional distinction is the narrower ensemble dispersion produced by FAST-ML, suggesting that the hybrid framework can maintain forecast accuracy while reducing unnecessary spread under relatively well-constrained environmental conditions.

\begin{figure}[H]
    \centering
    \includegraphics[width=0.95\textwidth]{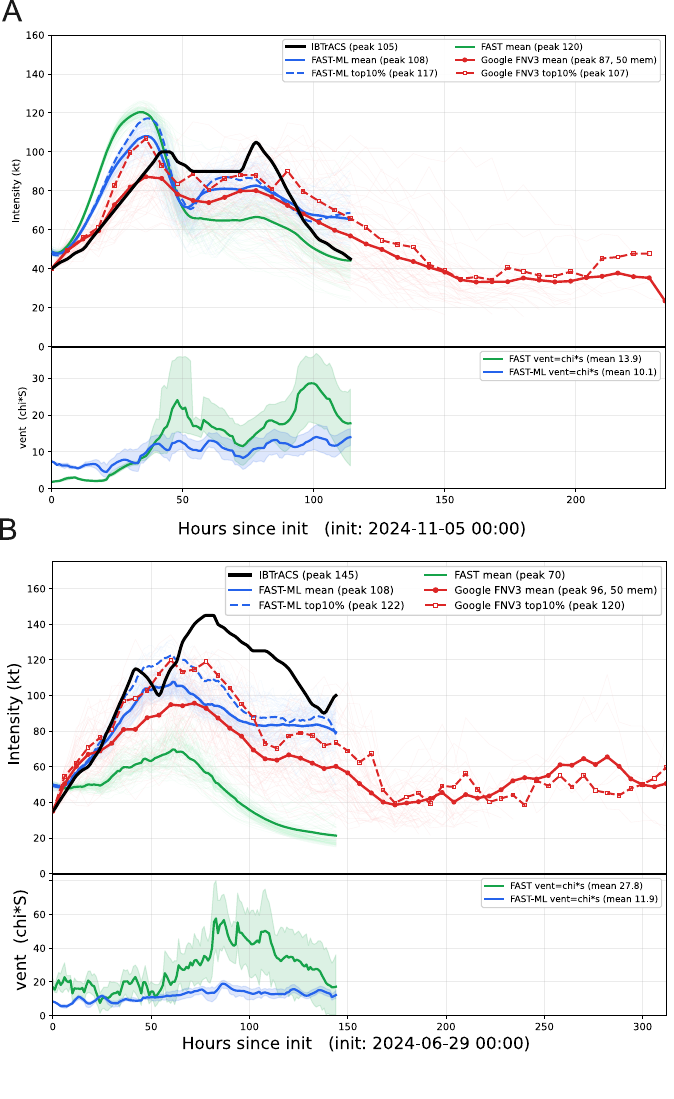}
    \caption{FAST-ML achieves comparable or better intensity-forecast skill than FNV3 for the selected Atlantic cases. Ensemble intensity predictions for (A) Hurricane Rafael (2024) and (B) Hurricane Beryl (2024). In these cases, FAST-ML (blue) produces ensemble-mean peak intensities closer to the observed peaks than Google FNV3 (red), although Beryl's peak remains underestimated.}
    \label{fig:atlantic_ensemble_outperform}
\end{figure}

\begin{figure}[H]
    \centering
    \includegraphics[width=0.95\textwidth]{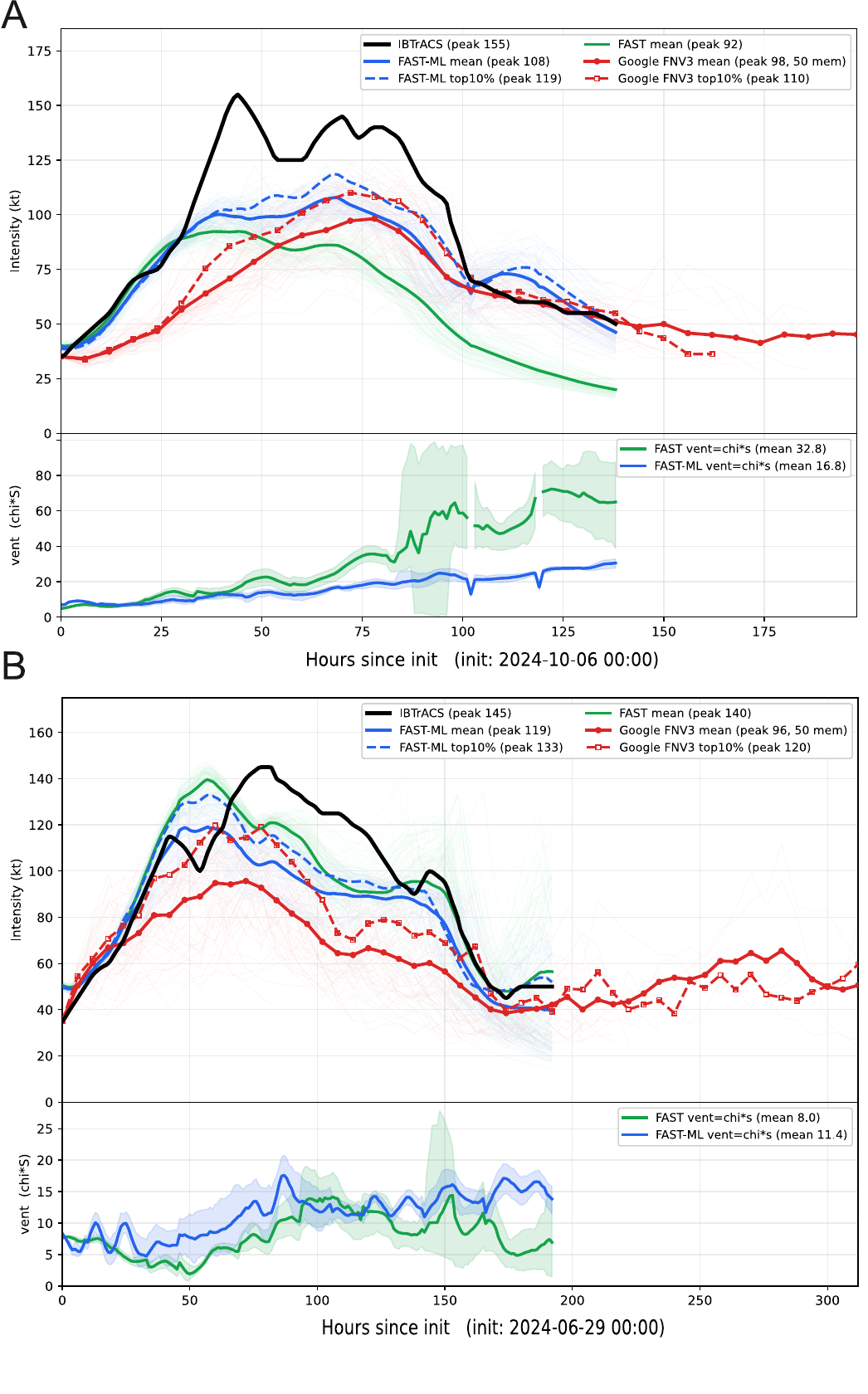}
    \caption{
Probabilistic intensity forecasting driven by Google FNV3 track ensembles for extreme Atlantic RI cases. Ensemble intensity forecasts for (A) Hurricane Milton (2024) and (B) Hurricane Beryl (2024). Top panels show the forecast intensity trajectories from FAST-ML (blue), Google FNV3 (red), and IBTrACS observations (black). For Hurricane Milton (2024), bottom panels display the temporal evolution of the diagnosed FAST-ML ventilation parameter ($\chi S$), showing a lower level than FAST that facilitates RI.}
    \label{google_tracks}
\end{figure}

For the selected Rafael and Beryl cases, FAST-ML more closely reproduces aspects of the observed intensity evolution than FNV3 under the evaluated diagnostic configuration. For Hurricane Rafael (2024)(Fig.~\ref{fig:atlantic_ensemble_outperform}A), the FNV3 ensemble mean reaches only 87 kt compared to the observed peak of 105 kt, whereas FAST-ML attains 108 kt and reproduces the observed intensification trajectory more closely. A similar pattern is evident for Hurricane Beryl (2024) (Fig.~\ref{fig:atlantic_ensemble_outperform}B) under the baseline diagnostic configuration. Although both systems underestimate the observed Category-5 peak intensity of 145 kt, FAST-ML captures a substantially larger fraction of the observed RI than FNV3. 
The alternative FNV3-track configuration examined below (Fig.~\ref{google_tracks}B) increases the FAST-ML ensemble-mean peak from 108 to 119~kt, although both configurations underestimate the observed peak of 145~kt.

\begin{figure}[H]
    \centering
    \includegraphics[width=0.95\textwidth]{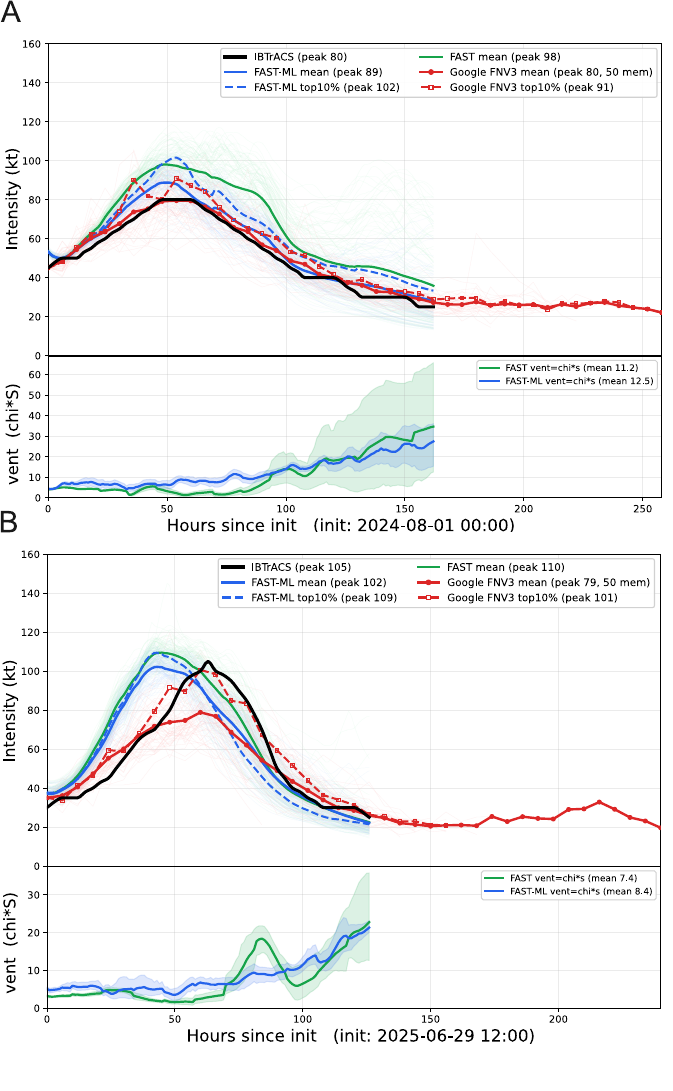}
    \caption{Zero-shot cross-basin generalization in the Eastern Pacific (EP). Ensemble predictions for (A) Hurricane Carlotta (2024) (Init: 2024-08-01, peak 80~kt) and (B) Hurricane Flossie (2025) (Init: 2025-06-29, peak 105~kt).}
    \label{fig:ep_zero_shot}
\end{figure}

To further assess model generalization, we also performed a zero-shot experiment using Eastern Pacific TCs despite training exclusively on North Atlantic storms. This experiment provides a stringent test of whether the trained FAST-ML model captures transferable physical relationships rather than basin-specific statistical characteristics. 
For Hurricane Carlotta (2024)(Fig.~\ref{fig:ep_zero_shot}A), FAST-ML predicts a peak intensity of approximately 89 kt, compared with the observed peak of 80 kt. 
Consistent performance is also observed for Hurricane Flossie (2025) (Fig.~\ref{fig:ep_zero_shot}B), where FAST-ML predicts a peak intensity of 102 kt compared to the observed 105 kt, whereas FNV3 reaches only 79 kt.
These results suggest that the environmental controls learned by the hybrid physical framework retain predictive relevance outside the basin used for training. More broadly, the ability to maintain forecast skill in an unseen basin indicates that the FAST-ML is learning physically meaningful storm--environment relationships that are transferable across different TC climatologies, rather than relying primarily on basin-specific statistical patterns.

\begin{figure}[H]
    \centering
    \includegraphics[width=0.95\textwidth]{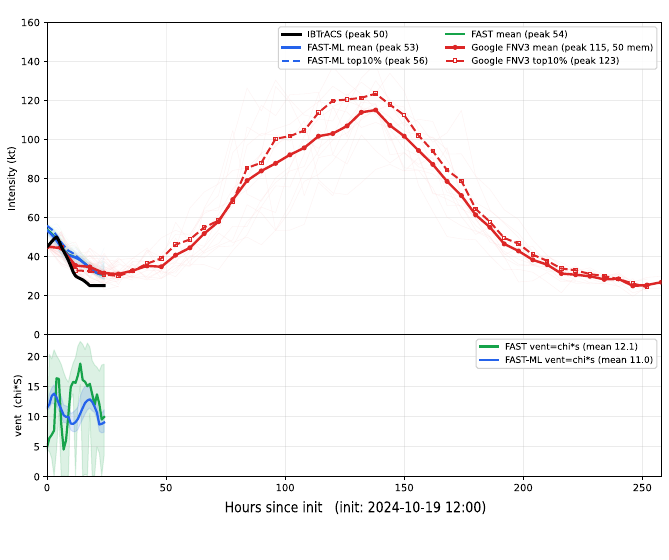}
    \caption{Mitigation of catastrophic false alarms via governing physical equations. Ensemble intensity forecasts for Hurricane Nadine (2024). Time series of forecast intensity from FAST-ML (blue), Google FNV3 (red), and IBTrACS observations (black). The lower panel shows the evolution of the diagnosed ventilation parameter ($\chi S$) from FAST-ML.}
    \label{fig:false_alarm_mitigation}
\end{figure}

Beyond raw intensity accuracy, a model's ability to avoid spurious over-forecasts of extreme intensity is critical for operational trust, since false alarms of major-hurricane strength can trigger unnecessary evacuations and erode confidence in future warnings. A particularly informative comparison emerges from a false-alarm scenario. Tropical Storm Nadine (Fig.~\ref{fig:false_alarm_mitigation}) reached a maximum observed intensity of only 50 kt. FNV3, however, predicts a substantially stronger storm, with an ensemble-mean peak approaching 115 kt. FAST-ML instead predicts a peak intensity of 53 kt, closely matching observations. Inspection of the diagnosed ventilation evolution provides insight into the source of this discrepancy. FAST-ML identifies a highly ventilated environment unfavorable for sustained intensification, and because the ventilation term enters the governing equations directly, the physical model constrains cyclone growth despite favorable conditions implied by some ensemble members. This example highlights a key distinction between the two modeling philosophies. In FAST-ML, environmental regulation is embedded directly within the governing dynamics, whereas in purely data-driven systems the suppression of extreme false alarms must be learned implicitly from historical examples.

These selected case studies show that FAST-ML can achieve intensity-forecast performance comparable to or better than FNV3 under the evaluated input configurations. More importantly, the framework provides physically interpretable pathways connecting environmental conditions to forecast outcomes, allowing forecast successes and failures to be traced directly to diagnosed environmental controls. The results suggest that embedding machine learning within a physically constrained dynamical framework offers a viable alternative to purely data-driven intensity prediction, particularly in situations where physical consistency, cross-basin generalization, and computational efficiency are important.

To evaluate the compatibility of FAST-ML with external AI-generated track guidance, we investigate its intensity predictions when driven by FNV3 track ensembles. Shared track guidance controls for differences in predicted storm trajectories, although the use of ERA5 environmental fields by FAST-ML means that the two systems do not receive equivalent input information. Driving FAST-ML with identical AI-generated track ensembles serves a dual purpose: it leverages the highly accurate trajectory predictions produced by state-of-the-art global AI models, while simultaneously eliminating track forecast discrepancies to isolate and compare each system's intrinsic intensity modeling capability. Figure~\ref{google_tracks} illustrates probabilistic predictions for Hurricane Milton (2024) and Hurricane Beryl (2024) under this controlled configuration.

For Hurricane Milton (Fig.~\ref{google_tracks}A), FAST-ML driven by FNV3 track guidance produces an ensemble-mean peak intensity of 108~kt, compared with 98~kt for FNV3, 92~kt for baseline FAST, and an observed peak of 155~kt. FAST-ML therefore reduces peak-intensity underestimation in this case, although the observed extreme remains substantially underestimated. The lower panel shows generally lower diagnosed ventilation in FAST-ML than in baseline FAST, consistent with reduced ventilation-related suppression and the higher predicted intensity. For Hurricane Beryl (Fig.~\ref{google_tracks}B), FAST-ML produces an ensemble-mean peak of 119~kt, compared with 96~kt for FNV3 and an observed peak of 145~kt. Comparison with the alternative track configuration in Figure~\ref{fig:atlantic_ensemble_outperform}B provides additional insight into the behavior of the learned closure. Across the two configurations, the baseline FAST peak changes from 70 to 140~kt, whereas the FAST-ML peak changes from 108 to 119~kt. Over the overlapping forecast period, the baseline ventilation diagnostics also differ markedly between configurations, while the FAST-ML diagnostics remain within a more similar range. These patterns suggest lower sensitivity of the learned closure and predicted peak intensity to the track configurations examined in this case. Baseline FAST nevertheless predicts a peak closer to observations under FNV3 track guidance, so the comparison does not establish uniformly better peak-intensity accuracy for FAST-ML. Its advantage is the combination of reduced under-intensification in the earlier configuration and more consistent behavior across the two configurations, while retaining higher ensemble-mean peak intensities than FNV3 in both. Together with the Rafael case, where FAST-ML moderates the excessive peak intensity predicted by baseline FAST, these results are consistent with an adaptive environmental closure that can mitigate both excessive and insufficient intensification. These case studies support the potential of coupling FAST-ML with external track guidance while retaining interpretable ventilation diagnostics. The comparison with FNV3 remains conditional on the evaluated inputs, since FAST-ML uses ERA5 environmental fields in these diagnostic experiments.

\begin{figure}[H]
    \centering
    \includegraphics[width=0.95\textwidth]{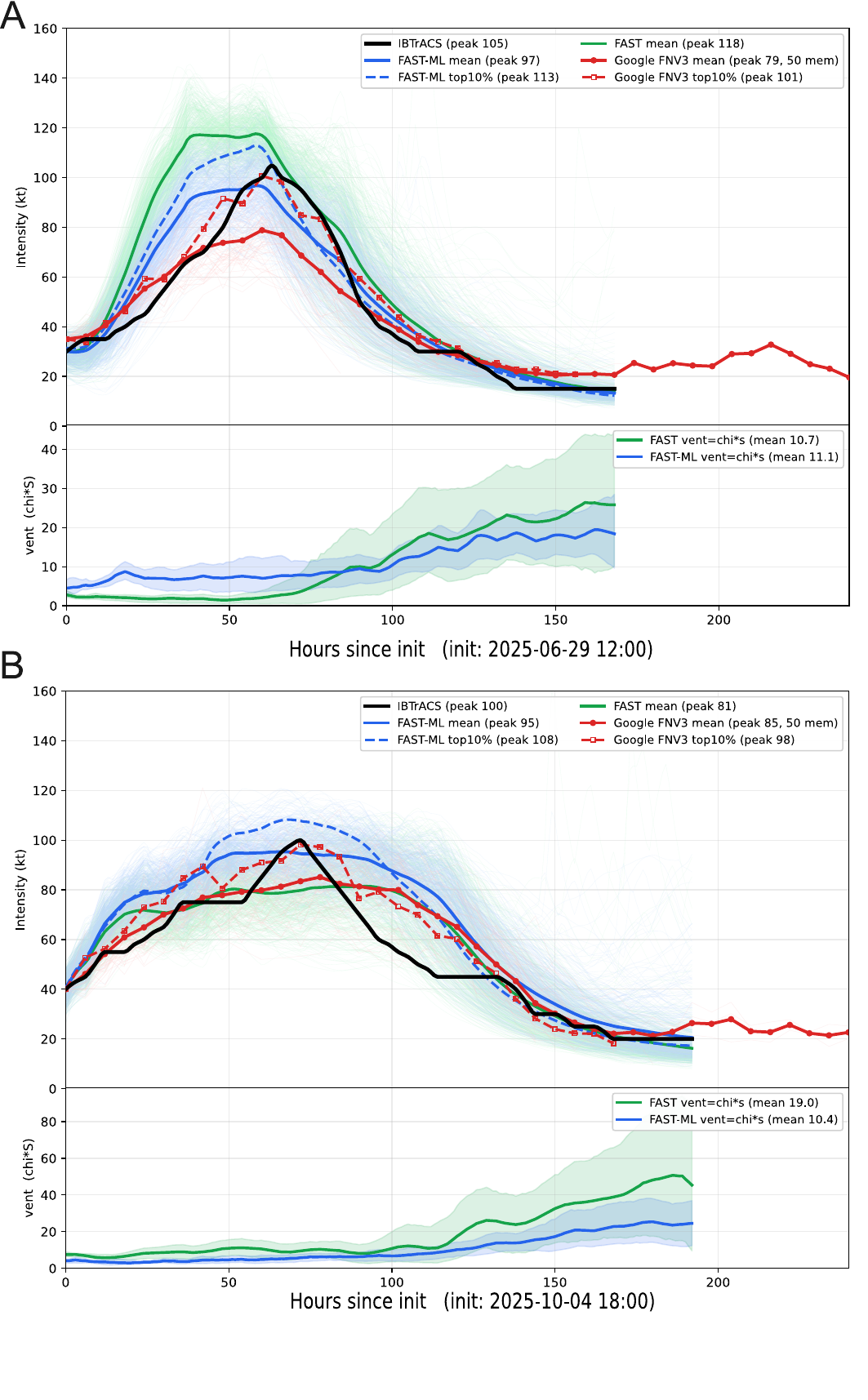}
\caption{Operational forecasting case studies using GEFS forecast fields and 1,000 track realizations. Ensemble intensity predictions for (A) Hurricane Flossie (2025) and (B) Hurricane Priscilla (2025). FAST-ML is driven by operational GEFS environmental forecast fields and 1,000 track realizations without reanalysis inputs. FAST-ML (blue) demonstrates competitive intensity-forecast performance relative to Google FNV3 (red) in these two cases.}
    \label{GEFS Forecasting}
\end{figure}

Moving beyond reanalysis-driven diagnostic experiments, we assess FAST-ML in a fully forecast-driven configuration using operational GEFS environmental fields and tracks (Fig.~\ref{GEFS Forecasting}). FAST-ML uses 1,000 synthetic track realizations generated from GEFS track distributions for Hurricane Flossie (2025) (Fig.~\ref{GEFS Forecasting}A) and Hurricane Priscilla (2025) (Fig.~\ref{GEFS Forecasting}B). In these selected cases, FAST-ML produces intensity forecasts competitive with FNV3 despite errors in the forecast environmental fields and tracks. These results demonstrate the feasibility of coupling FAST-ML with operational forecast inputs, although broader evaluation is needed to establish the consistency of its operational performance.

\subsection{Spatial Probabilistic Track and Wind Hazard Forecasting}

\begin{figure}[H]
    \centering
    \includegraphics[width=0.8\textwidth]{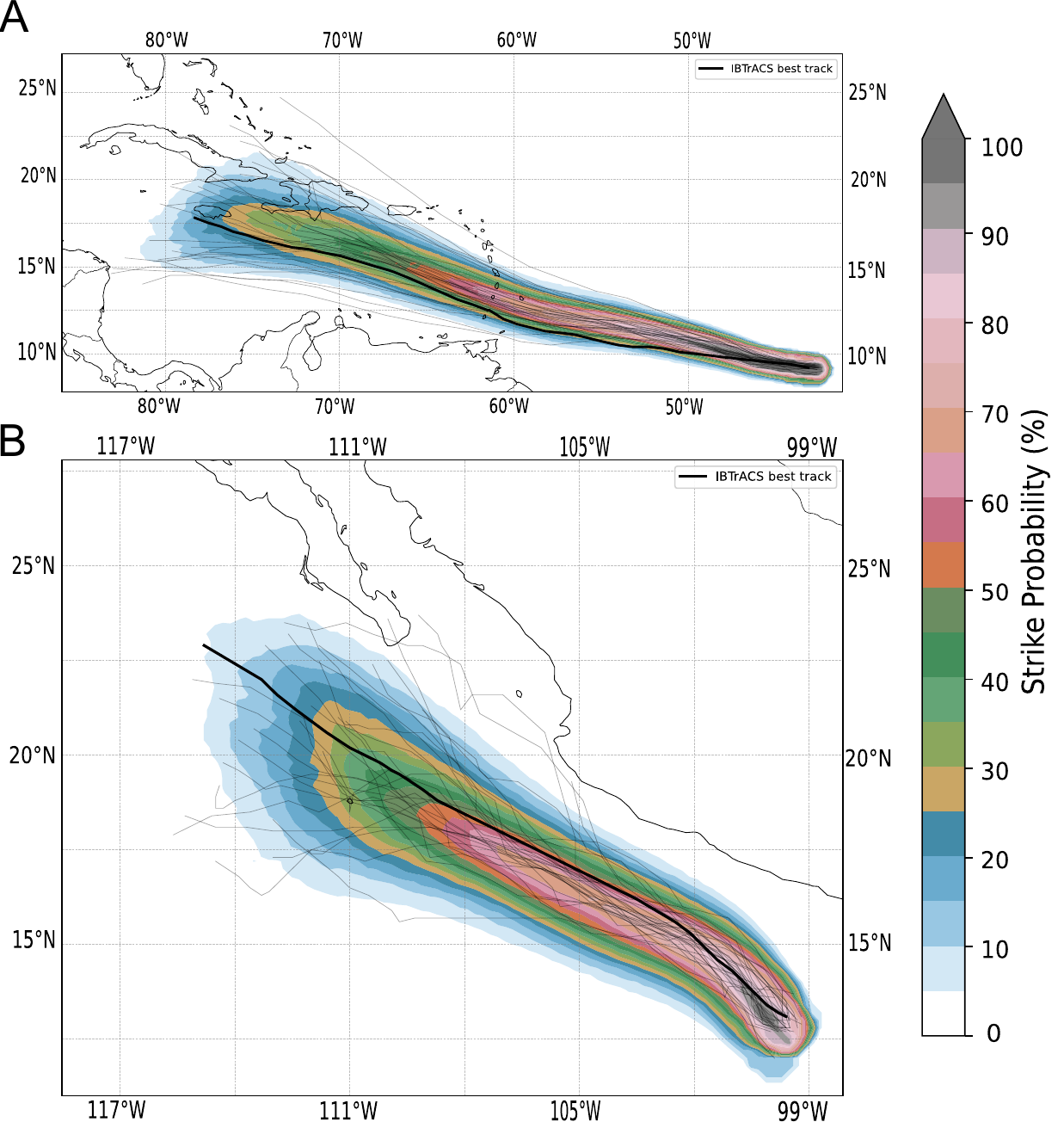}
    \caption{
Spatial track-strike probability forecasts driven by GEFS for
(A) Hurricane Beryl (2024) and
(B) Hurricane Flossie (2025).
Gray curves denote individual ensemble track realizations, the black curve denotes
the observed IBTrACS best track, and shaded contours show the probability that the
TC center passes within 75~km of each location at any time during the forecast
period, as defined in Eq.~\ref{eq:strike_probability}.}
    \label{fig:track_strike}
\end{figure}

The fully forecast-driven GEFS configuration further enables FAST-ML ensemble
predictions to be translated into spatially resolved probabilistic hazard products.
We demonstrate this capability using Hurricane Beryl (2024) in the North Atlantic
and Hurricane Flossie (2025) in the Eastern Pacific. These two cases provide
complementary examples of how uncertainty in storm trajectory and intensity can be
propagated from individual ensemble forecasts into spatial track-strike and surface-wind
exceedance probabilities.
Figure~\ref{fig:track_strike} shows the spatial track-strike probabilities defined
in Eq.~\ref{eq:strike_probability}. For Hurricane Beryl (2024), the ensemble
produces a coherent westward-propagating probability corridor extending from the
tropical Atlantic into the Caribbean Sea, with increasing cross-track dispersion
toward longer forecast lead times. Hurricane Flossie (2025) exhibits a concentrated
northwestward probability corridor offshore of southwestern Mexico. In both cases,
the observed IBTrACS best track remains largely within the principal ensemble
probability envelope, indicating that the GEFS-driven track ensemble captures the
dominant storm trajectory while retaining uncertainty in plausible alternative paths.

\begin{figure}[H]
    \centering
    \includegraphics[width=1.2\textwidth]{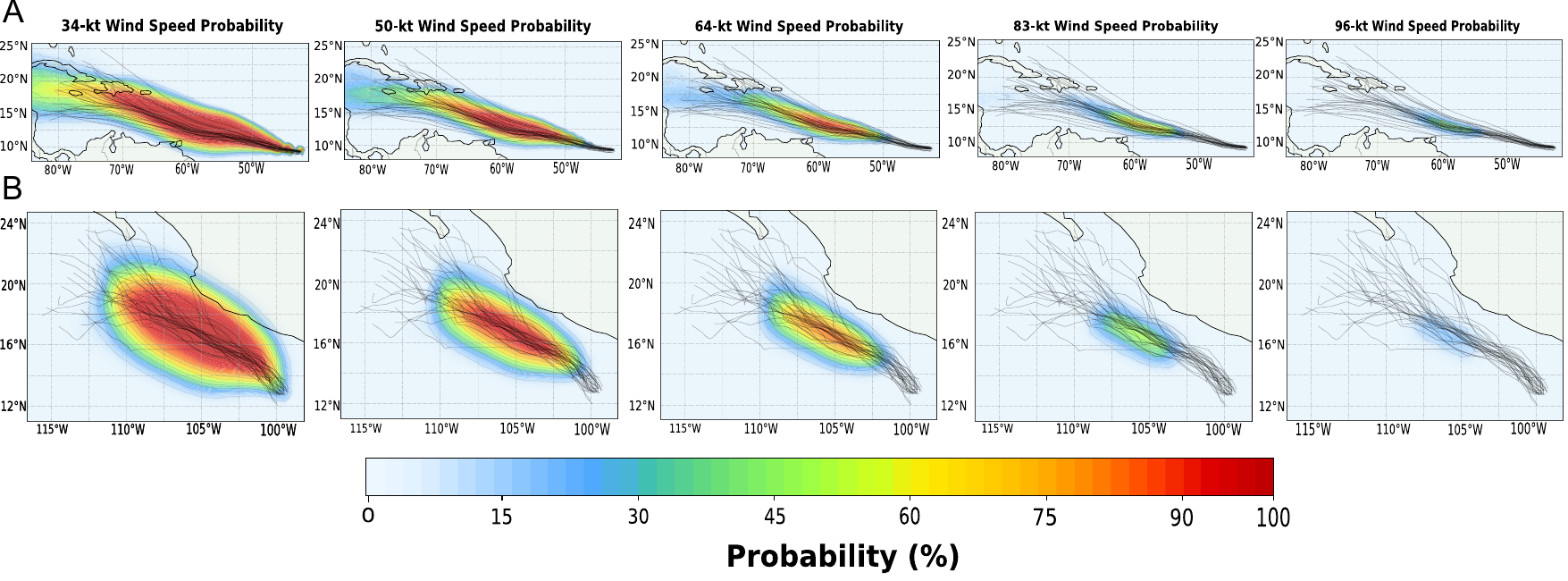}
    \caption{
Spatial wind-speed exceedance probabilities driven by GEFS for
(A) Hurricane Beryl (2024) and
(B) Hurricane Flossie (2025).
Columns show the probability that the maximum local surface wind speed during
the forecast period exceeds 34, 50, 64, 83, and 96~kt, respectively.
Gray curves denote the underlying ensemble track realizations, while shaded contours
represent the ensemble exceedance probability defined in
Eq.~\ref{eq:wind_exceedance_probability}. The progressive contraction of the
probability field with increasing threshold illustrates the spatial distribution of
increasingly severe forecast wind hazards.}
    \label{fig:wind_exceedance}
\end{figure}

Figure~\ref{fig:wind_exceedance} extends the analysis from storm-center trajectory
uncertainty to local surface-wind hazards using the exceedance probability defined
in Eq.~\ref{eq:wind_exceedance_probability}. The five columns correspond to
wind-speed thresholds of 34, 50, 64, 83, and 96~kt. For both storms, the spatial
extent and probability magnitude decrease systematically with increasing wind-speed
threshold, producing a nested hierarchy of increasingly severe wind-hazard regions.
For Hurricane Beryl (2024), high probabilities of tropical-storm-force winds extend
along a broad corridor from the tropical Atlantic into the Caribbean, whereas
hurricane-force and stronger wind probabilities become progressively concentrated
along the central portion of the forecast trajectory. The ensemble therefore captures
a broad region of elevated wind hazard while assigning increasingly smaller
probabilities to the most extreme local wind-speed thresholds.
For Hurricane Flossie (2025), the exceedance fields exhibit a similarly coherent
northwestward progression offshore of southwestern Mexico. High probabilities for
34--64~kt winds occupy a substantial portion of the primary forecast corridor, while
the 83- and 96-kt exceedance regions become progressively narrower and are confined
to the portion of the trajectory associated with the strongest forecast intensification.
Together, Figures~\ref{fig:track_strike} and~\ref{fig:wind_exceedance} demonstrate
that FAST-ML can propagate ensemble uncertainty beyond along-track intensity
prediction into spatially resolved hazard information. The track-strike product
quantifies where the TC center is likely to pass, whereas the wind-speed exceedance
product quantifies the probability that individual locations experience specified
levels of local surface wind. These complementary diagnostics provide a direct
connection between computationally efficient hybrid physics--ML forecasting and
probabilistic TC hazard assessment.

\subsection{Architectural Ablation and Component Analysis}

\begin{figure}[H]
    \centering
    \includegraphics[width=\textwidth]{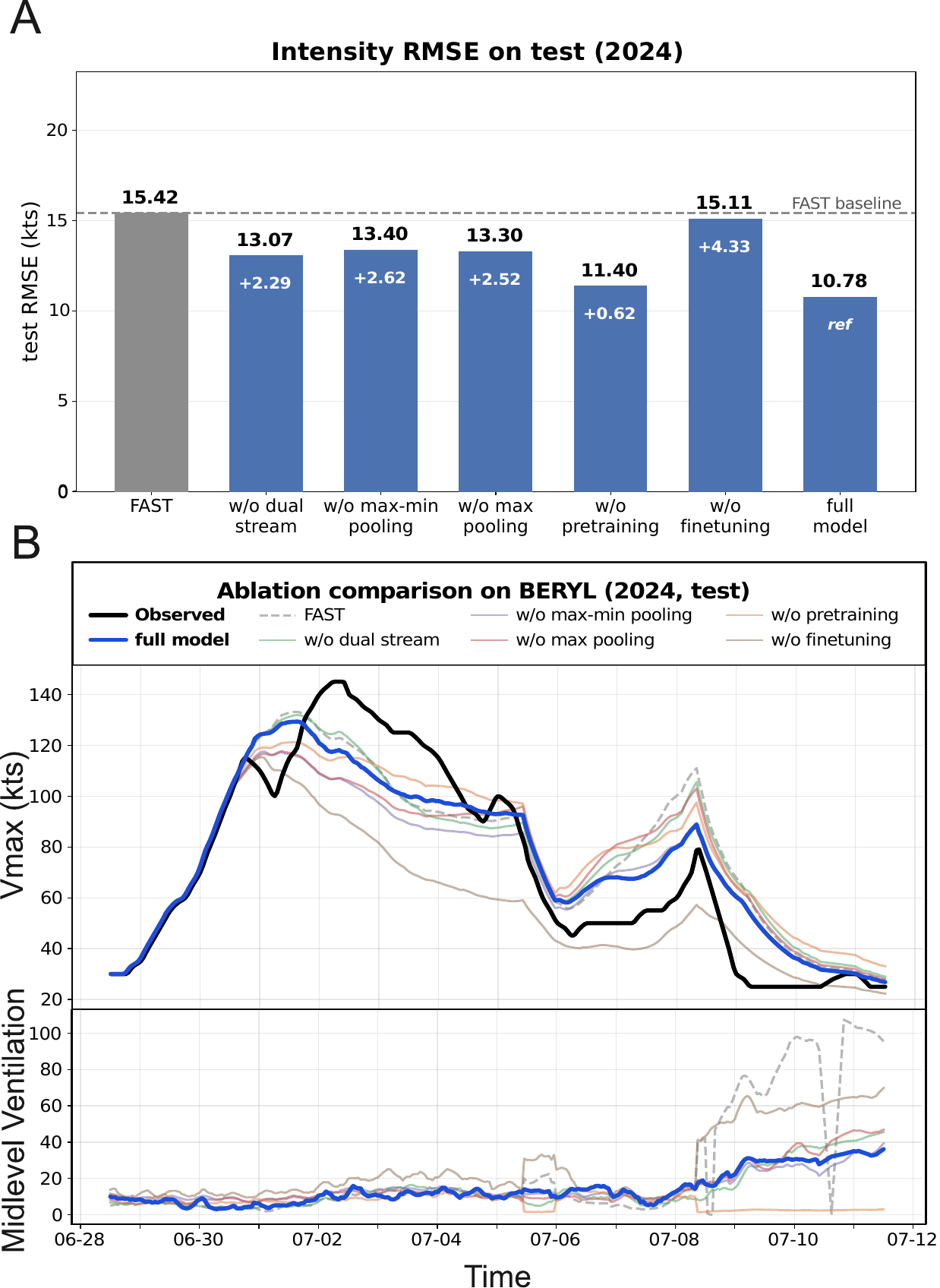} 
    \caption{Ablation study of the FAST-ML architecture on the 2024 test season. (A) Mean intensity RMSE across the 2024 test set for various architectural ablations compared to the baseline FAST model (dashed line) and the full FAST-ML framework (blue line). (B) Intensity trajectory of Hurricane Beryl (2024) highlighting the divergence of ablated variants during the RI phase.}
    \label{fig:ablation}
\end{figure}

The aggregate and case-study results in Sections~\ref{single track}--\ref{ensemble forecast} establish that FAST-ML outperforms the baseline FAST model, but they do not by themselves reveal which specific design choices are responsible for that improvement. To isolate these contributions, we compare the full FAST-ML model against five ablated variants on the independent 2024 test season (Figure~\ref{fig:ablation}A). The full model achieves a mean test RMSE of 10.78 kt, compared with 15.42 kt for the baseline FAST model. These experiments provide insight into whether the improvement arises primarily from the neural representation itself, the training strategy, or the physically motivated architectural constraints imposed on the closure.
The largest degradation occurs when the end-to-end differentiable fine-tuning stage is removed. In this configuration, the FAST-ML is trained only through physics-guided pretraining to reproduce the analytical FAST closure, but is not subsequently optimized against observed intensity evolution. This ablated model produces a test RMSE of 15.11 kt. This result indicates that simply replacing the analytical closure with a neural approximation of the same parameterization does not improve forecast skill. Instead, the performance gain arises from allowing the closure to depart from the original analytical formulas during observation-driven optimization, while still remaining embedded within the physical solver.

The architectural structure of the FAST-ML also contributes substantially to model skill. Removing the dual-stream separation increases RMSE to 13.07 kt, while eliminating Max-Min pooling increases RMSE to 13.40 kt, and removing maximum pooling entirely further degrades the performance to 13.30 kt. These degradations indicate that the model benefits from explicitly separating kinematic and thermodynamic components of environmental forcing and from retaining localized spatial extremes in addition to domain-mean environmental information. The importance of these design choices is physically consistent with the ventilation problem: storm intensity is influenced both by large-scale synoptic conditions and by localized asymmetric structures, such as dry-air intrusions and shear-related anomalies, that may be smoothed out by average pooling alone.
Removing physics-guided pretraining also degrades performance, increasing RMSE to 11.40 kt. Although the effect is smaller than removing end-to-end fine-tuning, it remains substantial. This suggests that pretraining provides a useful physically informed initialization, constraining the model toward stable regions of closure-parameter space before full observation-driven optimization. The two-stage training procedure therefore plays a complementary role: pretraining anchors the FAST-ML to physically meaningful behavior, while differentiable fine-tuning allows it to correct deficiencies in the original analytical parameterization.

Unlike the ensemble-mean comparison in Fig.~\ref{fig:atlantic_ensemble_outperform}B, panel B here shows the deterministic (single-trajectory) forecast and illustrates these dependencies for Hurricane Beryl (2024). Ablated variants lacking either end-to-end fine-tuning or the dual-stream design fail to reproduce the RI phase and instead converge toward trajectories closer to the baseline FAST simulation. This case-level behavior is consistent with the aggregate RMSE results and reinforces the interpretation that both the architectural inductive biases and the training curriculum are necessary for robust environmental parameterization. Overall, the ablation experiments demonstrate that FAST-ML's performance is not attributable to a generic increase in model flexibility alone, but to the combination of physically structured inputs, closure-specific architecture, and end-to-end coupling with the differentiable physical solver.

\subsection{Computational Efficiency and Operational Flexibility}

A central motivation for embedding machine learning within a reduced-order physical model is to improve forecast skill without sacrificing the computational efficiency that makes FAST attractive for large-ensemble applications. Because the FAST-ML replaces only two scalar environmental parameterizations rather than generating the full atmospheric state, its computational overhead remains minimal. For a 145-hour forecast, neural-network inference requires 0.42 s per ensemble member, compared with 18.7 s for the Python-based ODE integration itself, corresponding to an overhead of approximately 2\%. A 100-member dual-mode ensemble, corresponding to 200 total simulations, completes in under 3 wall-clock minutes on a single computing node with 32 CPU cores and one NVIDIA A100 GPU. Training converges in under 2 GPU-days on a single NVIDIA A100 GPU.

By comparison, FNV3 requires roughly 60 s per member on a TPU v5p and an estimated 1,960 TPU-days for training (Table~\ref{tab:resource_comparison}). This comparison should be interpreted carefully because the two systems solve different prediction problems. FNV3 generates full three-dimensional global atmospheric fields, whereas FAST-ML is specialized for one-dimensional along-track TC intensity forecasting. Thus, the lower computational cost of FAST-ML does not imply that it is a general substitute for global AI weather models. Rather, it indicates that for the specific task of probabilistic TC intensity forecasting, a physics-constrained reduced-order framework can achieve useful skill with substantially smaller computational requirements.
This efficiency has important implications for ensemble design. Because FAST-ML is track agnostic, externally generated track ensembles can be resampled, or updated and then paired with rapidly generated intensity trajectories without rerunning a full global atmospheric forecast. This makes it possible to construct large probabilistic intensity ensembles conditioned on alternative track hypotheses, updated track guidance, or targeted uncertainty scenarios. In contrast, models that generate full global atmospheric states are computationally less flexible when many alternative track realizations or sensitivity experiments must be evaluated.
The computational profile of FAST-ML therefore supports a complementary role for hybrid reduced-order models within the broader forecasting ecosystem. Global AI and NWP systems provide dynamically rich atmospheric forecasts, while FAST-ML can use externally supplied tracks and environmental fields to generate rapid, physically constrained intensity ensembles at low marginal cost. This design is especially useful for high-impact forecasting contexts in which many plausible track--intensity combinations must be evaluated quickly. More broadly, these results illustrate how machine learning can be embedded within physically based reduced-order models to improve predictive skill without sacrificing interpretability, scalability, or operational practicality, providing a potential template for other Earth-system forecasting applications.

\begin{table}[htbp]
\centering
\small
\caption{Resource comparison between FAST-ML and Google FNV3. FNV3 metrics are cited from published benchmarks. Note: FAST-ML is specialized for 1D along-track intensity, whereas FNV3 generates full 3D global atmospheric fields. These resource metrics are not intended as a direct efficiency comparison owing to differences in forecasting scope and hardware. The GEFS-driven FAST-ML forecasting configuration uses 1,000 ensemble members and can be readily scaled to larger ensembles.}
\label{tab:resource_comparison}
\begin{tabular}{l cc}
\toprule
\textbf{Metric} & \textbf{FAST-ML (Ours)} & \textbf{Google FNV3} \\
\midrule
Target Scope          & TC Intensity (1D)        & Global Atmosphere (3D) \\
Neural Parameters     & $\sim 1.2$ M             & $> 100$ M              \\
Memory Footprint      & $< 2$ GB (VRAM)          & $\ge 16$ GB (HBM)      \\
Training Compute      & $< 2$ GPU-days           & $\sim 1,960$ TPU-days  \\
\midrule
Single-Member Time & $\sim 19$ s (145-hour forecast) & $\sim 60$ s (15-day forecast) \\
Inference Hardware    & 1 CPU Core               & 1 TPU v5p              \\
Relative FLOPs        & $\sim 10^{-5}$           & 1.0 (Baseline)         \\
\midrule
Ensemble Configuration & 100 members per mode (200 simulations total) & 50 members \\
Ensemble Wall-Clock   & $< 3$ min                & $\sim 1$ min           \\
Deployment Scale      & 1 Node (32 CPU, 1 GPU)   & 50 TPU v5p             \\
\bottomrule
\end{tabular}
\end{table}

\newpage
\subsection{Model Interpretability and Physical Alignment}
The FAST governing equations represent environmental suppression of TC intensity through the combined ventilation product $\chi S$, where $S$ denotes the kinematic forcing associated with environmental vertical wind shear and $\chi$ represents the thermodynamic susceptibility of the storm to ventilation of low-entropy environmental air. In the original physical formulation, $S$ is expected to reflect the lower-to-upper-tropospheric shear structure, while $\chi$ is closely related to the mid-level entropy deficit that regulates ventilation-induced suppression. Because FAST-ML estimates both quantities directly from multi-dimensional environmental fields while learning the relative contributions of the prescribed input variables, pressure levels, and spatial locations, it is crucial to verify that the learned closure recovers physically meaningful environmental controls rather than relying on spurious statistical correlations.

\begin{figure}[H]
    \centering
    \includegraphics[width=1.0\textwidth]{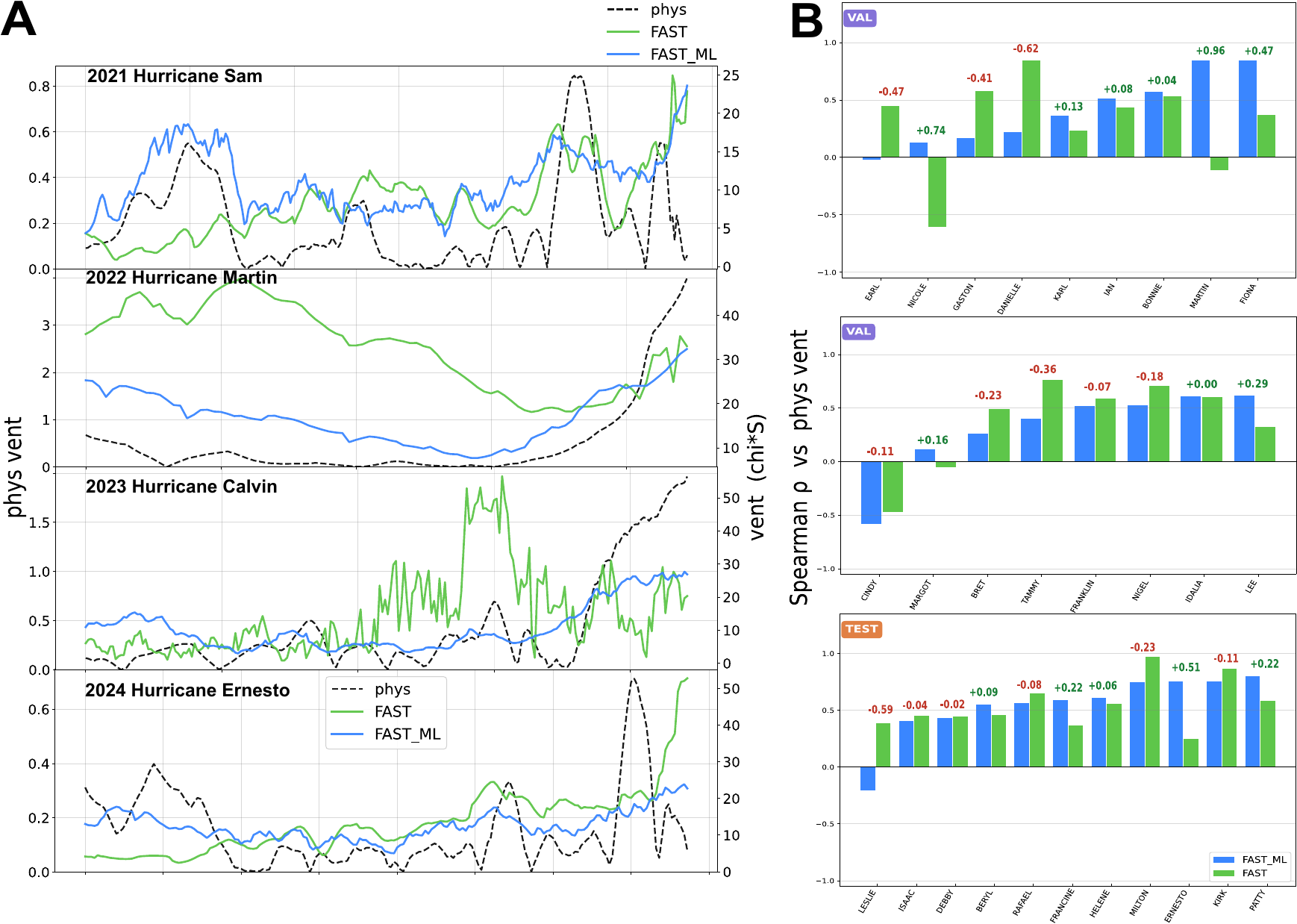}
    \caption{Physical Validation of the Learned Ventilation Closure. (A) Time series comparing the physical vertically integrated eddy entropy flux (dashed black, $F_{vent}$), the baseline FAST empirical ventilation (green), and the FAST-ML neural ventilation (blue) for selected storms. (B) Spearman correlation coefficients ($\rho$) between predicted ventilation ($\chi S$) and physical ventilation across the validation and 2024 test sets. FAST-ML shows stronger agreement with the physical diagnostic for some storms, with performance varying across cases.}
    \label{fig:phys_vent_correlation}
\end{figure}

Building upon the diagnostic benchmark established in
Eqs.~\ref{eq:physical_ventilation_index} and \ref{eq:eddy_entropy_flux},
Figure~\ref{fig:phys_vent_correlation} evaluates the physical alignment
of the learned ventilation closure. The time-series comparisons
(Fig.~\ref{fig:phys_vent_correlation}A) show that FAST-ML captures
aspects of the temporal variability in the independently computed
eddy entropy-flux diagnostic. Across the validation and independent
2024 test sets (Fig.~\ref{fig:phys_vent_correlation}B), the Spearman
correlations indicate stronger agreement than baseline FAST for
some storms, although improvements are not uniform. These results
support the physical relevance of the learned closure while
highlighting storm-dependent differences in its agreement with
the diagnostic benchmark.

\begin{figure}[H]
    \centering
    \includegraphics[width=1.2\textwidth]{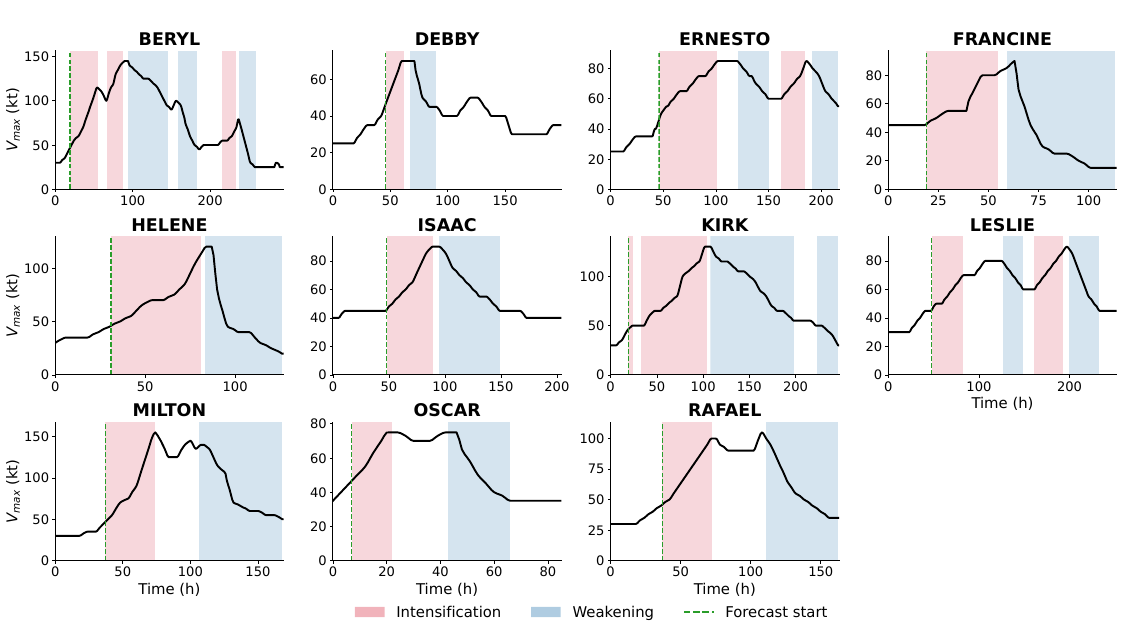}
    \caption{Storm Evolution and Phase Classification. Time series of maximum sustained wind speeds (Vmax, in knots) for storms in the 2024 independent test set. Forecast windows are segmented into discrete intensification (red shading) and weakening (blue shading) phases, providing the temporal basis for the phase-dependent saliency analyses.(Note: Hurricane Patty (2024) is excluded from the phase classification as it did not undergo RI.)}
    \label{fig:storm_phases}
\end{figure}

To capture these complex physical interactions, the model ingests atmospheric fields at multiple pressure levels—specifically atmospheric temperature ($T$), specific humidity ($Q$), zonal wind ($U$), meridional wind ($V$), and geopotential height ($Z$)—along with critical two-dimensional surface boundary fields: Sea Surface Temperature (SST) and Mean Sea Level Pressure (MSLP). 
To evaluate the physical consistency of the learned closure across these inputs, we compute gradient-based saliency with respect to both closure outputs, $\partial S/\partial x$ and $\partial \chi/\partial x$, for storms in the 2024 independent test set. Saliency measures sensitivity of the trained model to its inputs and should not be interpreted as a causal response of the physical system. The analysis is performed separately for intensification and weakening periods, as defined from each storm's forecast-window intensity evolution (Fig.~\ref{fig:storm_phases}). This phase-dependent analysis allows us to assess not only which atmospheric variables and pressure levels influence the learned closure, but also whether the model's sensitivity to environmental forcing changes as storms transition between growth and decay regimes. Because saliency analysis measures model sensitivity rather than causal response, these sensitivities should not be interpreted as establishing causal relationships between environmental conditions and storm intensity changes.

\begin{figure}[H]
    \centering
    \includegraphics[width=1.1\textwidth]{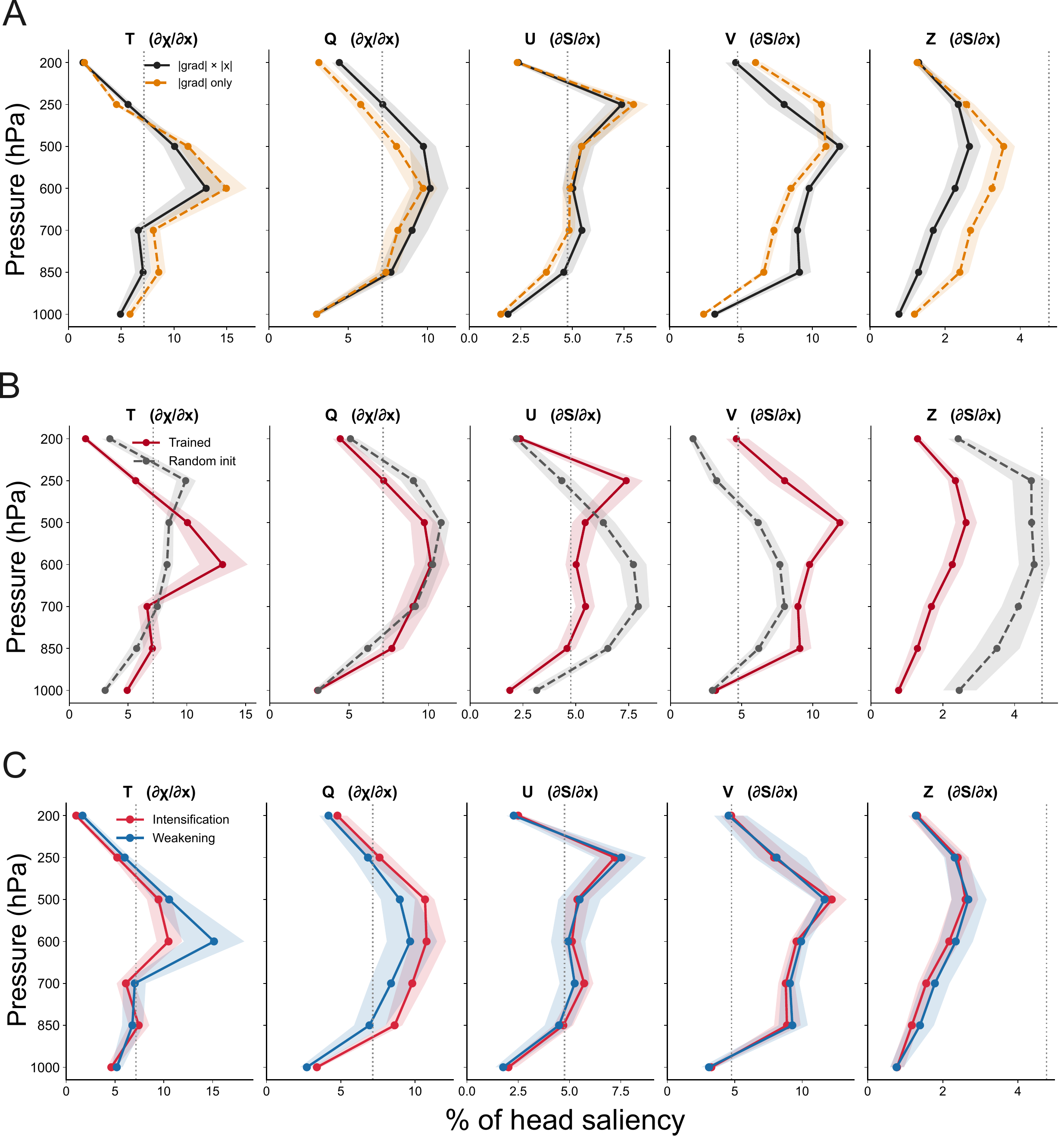}
    \caption{Vertical Distribution of Learned Saliency. Vertical profiles of relative feature importance (saliency) across pressure levels. (A) Comparison of saliency computation methods, showing that gradient-times-input ($|grad| \times |x|$) yields similar structures to pure gradients. (B) Comparison between the fully trained FAST-ML model (red) and a randomly initialized network (gray dashed), demonstrating that the physical structures are robustly learned during training. (C) Phase-dependent differences in vertical sensitivity between storm intensification (red) and weakening (blue) phases.}
    \label{fig:vert_saliency}
\end{figure}

As a preliminary robustness check, Figure~\ref{fig:vert_saliency}A compares two attribution metrics: raw gradients and gradient-times-input. The two approaches produce highly consistent vertical sensitivity structures, with dominant peaks appearing at similar pressure levels across variables. Although the relative magnitude of saliency differs modestly at individual levels, particularly for specific humidity in the mid-troposphere, the overall agreement indicates that the inferred vertical structures are not an artifact of a single attribution convention. A second validation is provided by comparison with a randomly initialized network of identical architecture (Figure~\ref{fig:vert_saliency}B). The random network exhibits weak and largely structureless vertical saliency, whereas the trained FAST-ML model develops coherent, physically localized sensitivity patterns. This contrast indicates that the structures discussed below are learned through training and are not simply inherited from the intrinsic variance or vertical scaling of the input fields.

\begin{figure}[H]
    \centering
    \includegraphics[width=1.15\textwidth]{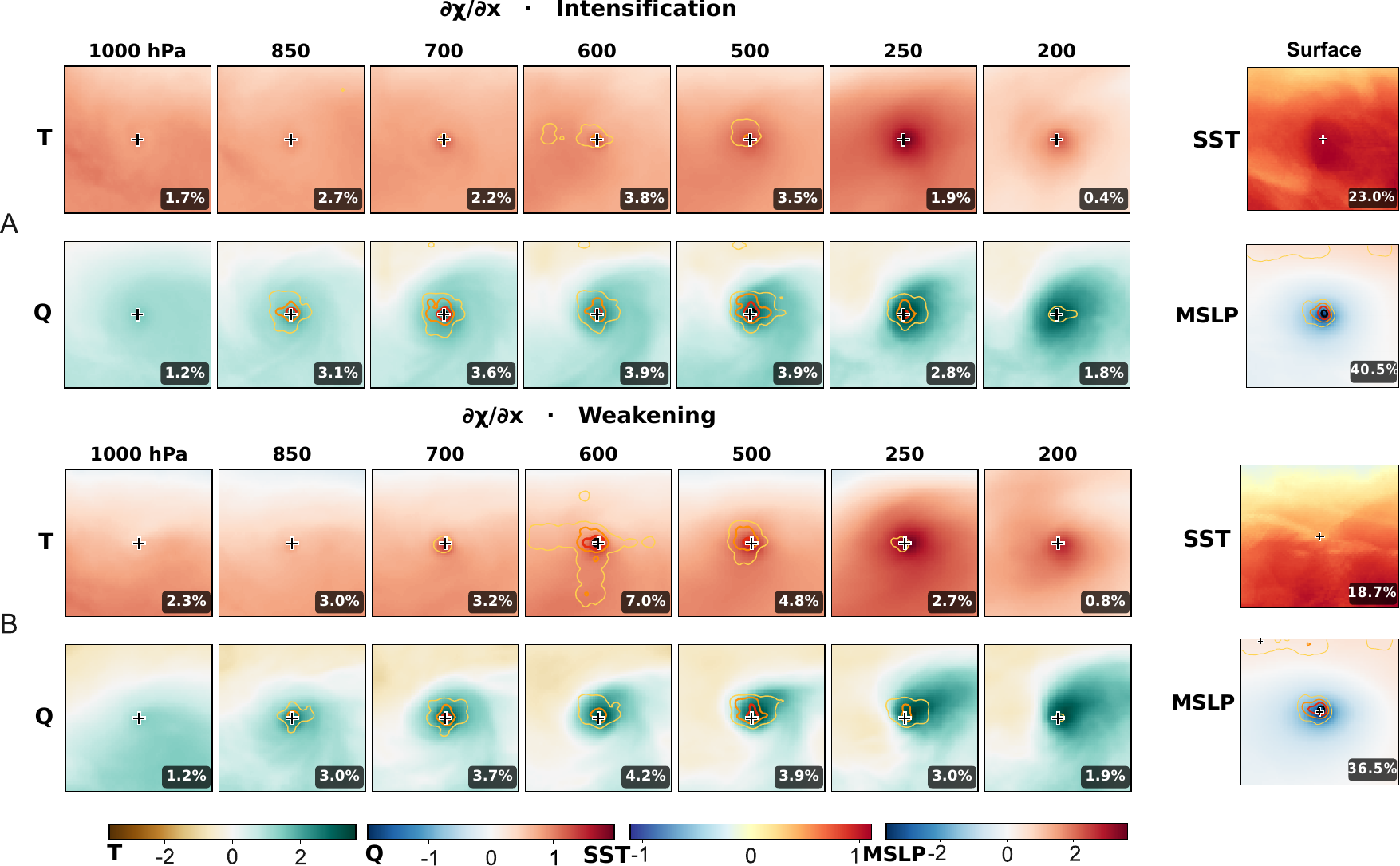}
    \caption{Composite Spatial Saliency Maps for Thermodynamic Closure ($\partial \chi / \partial x$). Composite spatial sensitivities of 3D atmospheric variables (Temperature $T$, Specific Humidity $Q$) alongside 2D surface boundary fields (Sea Surface Temperature SST, Mean Sea Level Pressure MSLP) averaged across all analyzed storms, partitioned by (A) Intensification and (B) Weakening phases. The model consistently targets mid-tropospheric levels (600--500~hPa) for thermodynamic corrections while capturing surface air-sea thermal coupling, aligning with physical ventilation mechanisms. (Note: Contours denote 30, 50, 70, and 90\% of the composite maximum saliency evaluated across the combined 2D/3D feature space; this convention applies to all subsequent spatial saliency figures. Visualized percentages in the figure represent total global importance partitioned across all 2D and 3D inputs, whereas numerical values cited in the text denote normalized relative importance evaluated within 3D atmospheric levels. )}
    \label{fig:composite_thermo}
\end{figure}

The thermodynamic closure exhibits a clear and physically interpretable vertical organization. As illustrated in the vertical saliency profiles (Figure~\ref{fig:vert_saliency}C), temperature and specific humidity sensitivities are heavily concentrated within the 500--600~hPa layer. This distinct mid-tropospheric maximum is further evident in the composite spatial saliency maps (Figure~\ref{fig:composite_thermo}). This structure is consistent with the entropy-deficit interpretation of $\chi$, in which ventilation is controlled by the flux of lower-entropy air from the middle troposphere into the storm inner core. Importantly, this vertical structure is not imposed on the neural network. Instead, the closure identifies the same layer emphasized by TC ventilation theory directly from the relationship between environmental fields, diagnosed closure parameters, and observed intensity evolution.

\begin{figure}[H]
    \centering
    \includegraphics[width=1.15\textwidth]{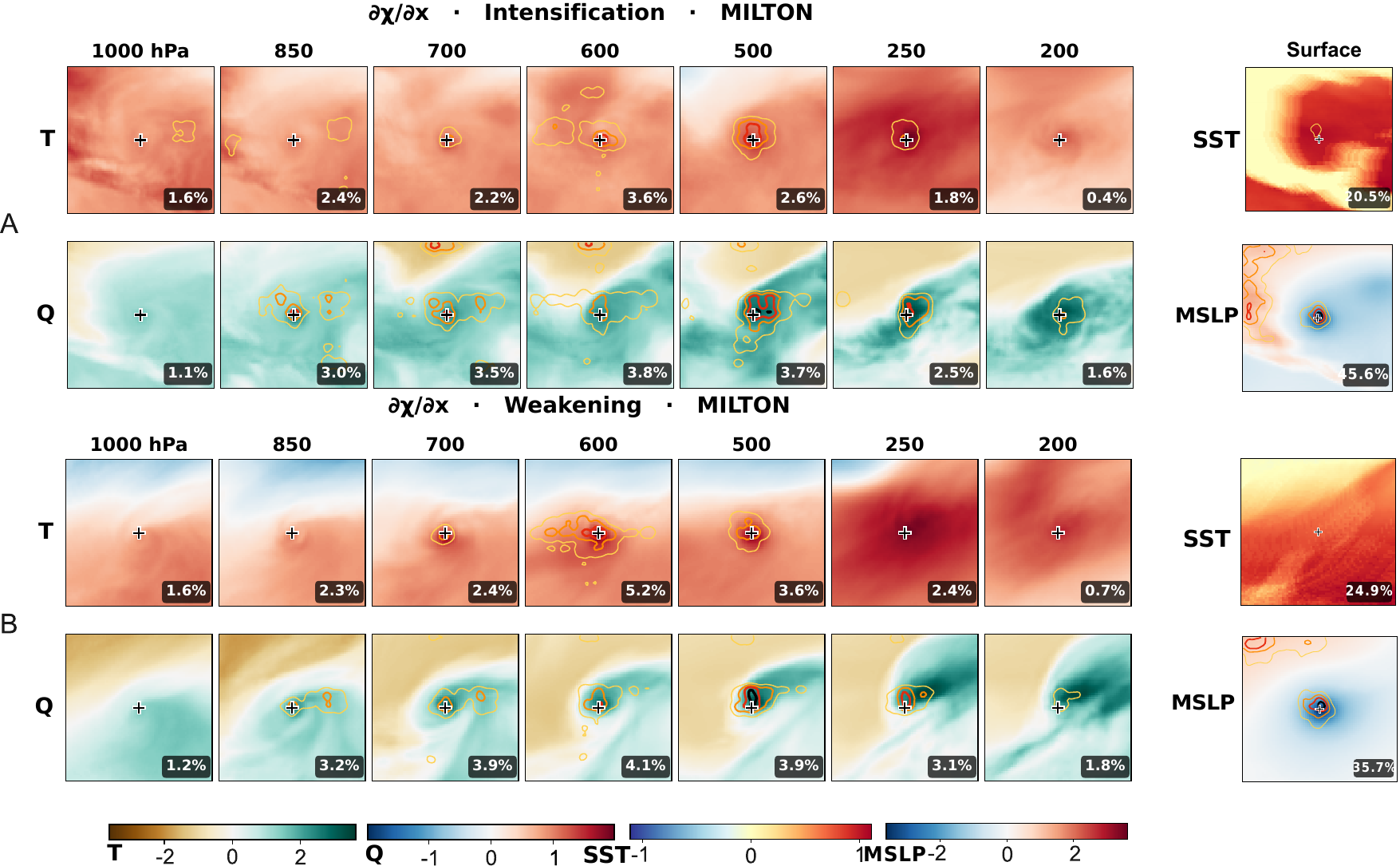}
    \caption{Thermodynamic and Surface Boundary Saliency for Hurricane Milton (2024). Spatial sensitivities of 3D atmospheric variables (Atmospheric Temperature $T$, Specific Humidity $Q$) alongside 2D surface boundary fields (Sea Surface Temperature SST, Mean Sea Level Pressure MSLP) for Milton during its (A) Intensification and (B) Weakening phases. During weakening, the model exhibits hyper-sensitivity to mid-level temperature anomalies (up to 13.1\% relative 3D importance at 600~hPa).}
    \label{fig:milton_thermo}
\end{figure}

The thermodynamic sensitivities also exhibit a notable dependence on storm phase and surface boundary coupling. To account for the dilution of absolute feature weights resulting from the inclusion of 2D surface fields (SST and MSLP), vertical saliency percentages reported below are evaluated as normalized relative importance within the 3D atmospheric variables, while the spatial maps visualize the global importance distribution across both 2D and 3D inputs.
In the composite analysis, normalized 600~hPa temperature sensitivity increases from 10.5\% during intensification to 15.1\% during weakening, whereas specific humidity sensitivity at the same level remains nearly unchanged or slightly decreases, from 10.8\% to 9.7\%. A similar pattern appears in the Hurricane Milton (2024) case study (Fig.~\ref{fig:milton_thermo}), where 600~hPa temperature sensitivity increases from 10.7\% during intensification to 13.1\% during weakening, while specific humidity sensitivity decreases modestly from 11.2\% to 10.4\%. 

The surface inputs also contribute to the learned thermodynamic closure. SST exhibits localized saliency during intensification, while the weakening-phase patterns emphasize mid-level temperature sensitivity and contributions from MSLP. These associations are compatible with the relevance of ocean surface conditions and atmospheric structure to TC evolution. However, saliency of SST or MSLP does not directly identify surface heat fluxes, establish causal mechanisms, or demonstrate an independently learned energy balance.

\begin{figure}[H]
    \centering
    \includegraphics[width=1.1\textwidth]{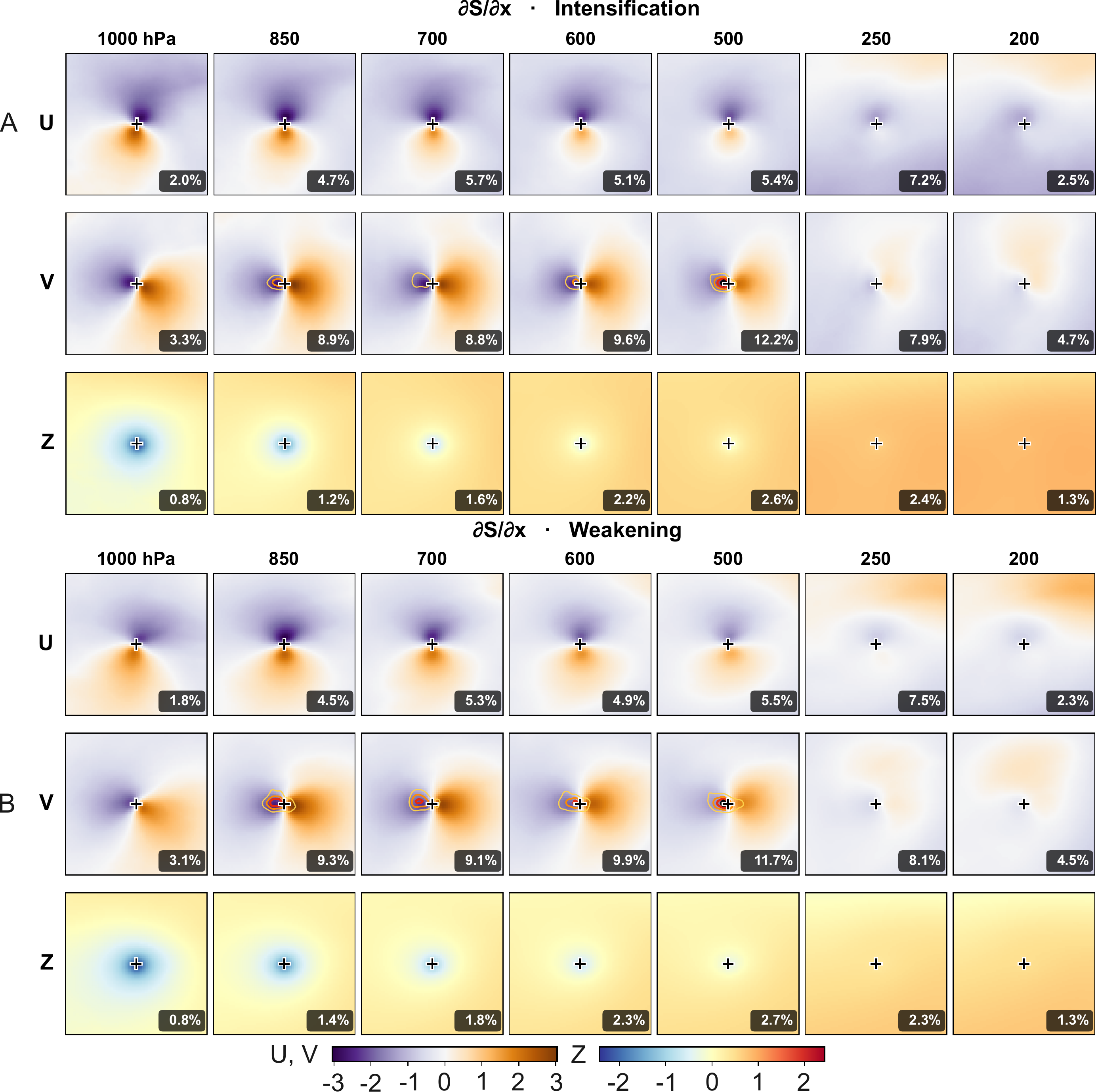}
    \caption{Composite Spatial Saliency Maps for Dynamic Tendency ($\partial S / \partial x$). Composite spatial sensitivities of Zonal Wind ($U$), Meridional Wind ($V$), and Geopotential Height ($Z$) across all analyzed storms, separated by \textbf{(A)} Intensification and \textbf{(B)} Weakening phases. Notably, the zonal wind ($U$) develops a secondary peak of importance in the upper troposphere (250 hPa), capturing the storm's outflow dynamics and upper-level shear constraints.}
    \label{fig:composite_kine}
\end{figure}

The kinematic closure also displays a different and more phase-invariant structure. Composite saliency maps for $\partial S/\partial x$ (Figure~\ref{fig:composite_kine}) show that meridional wind sensitivity is strongest in the lower-to-middle troposphere, with a maximum near 500~hPa of 12.2\% during intensification and 11.7\% during weakening. Zonal wind sensitivity exhibits a secondary maximum at 250~hPa, reaching 7.2\% during intensification and 7.5\% during weakening. This vertical pattern is consistent with the physical interpretation of $S$ as an effective shear-related quantity: lower- and mid-level winds contribute to the environmental steering and shear structure surrounding the vortex, while upper-tropospheric winds near 250~hPa capture the outflow-layer contribution to vertical shear. Geopotential height contributes comparatively weakly at all pressure levels, remaining below 3.2\%, suggesting that it primarily provides contextual information about the surrounding flow rather than acting as a dominant direct control on the diagnosed shear parameter.

\begin{figure}[H]
    \centering
    \includegraphics[width=1.1\textwidth]{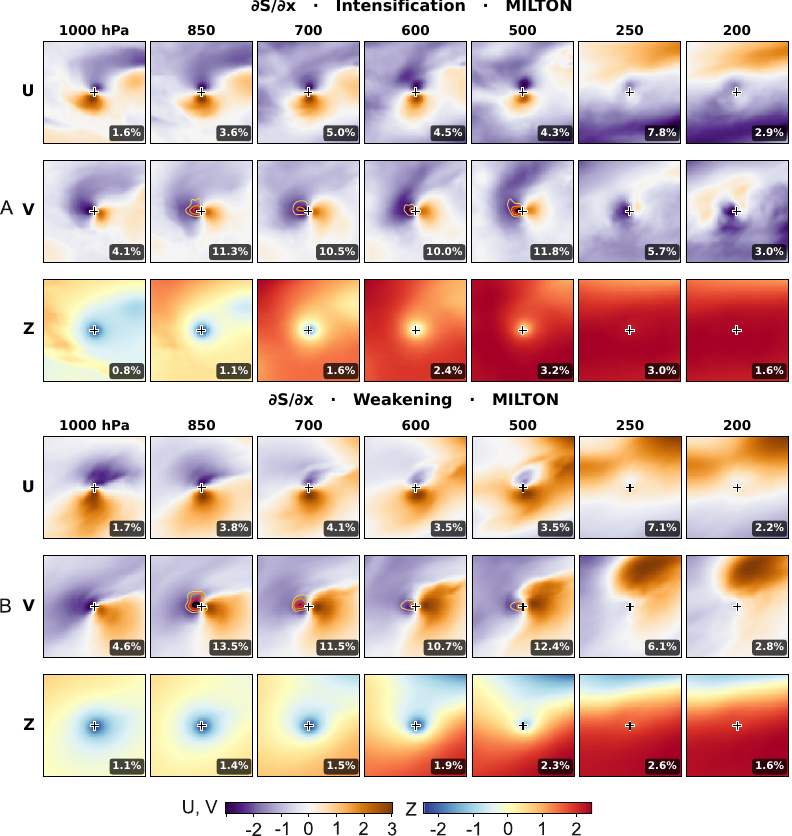}
    \caption{Wind Field Saliency for Hurricane Milton (2024). Specific spatial sensitivities of Zonal Wind ($U$), Meridional Wind ($V$), and Geopotential Height ($Z$) for Milton during its (A) Intensification and (B) Weakening phases. The wind field attention is appropriately partitioned between low-to-mid-level steering ($V$) and upper-level divergence ($U$ at 250 hPa).}
    \label{fig:milton_kine}
\end{figure}

The Hurricane Milton (2024) kinematic saliency maps (Figure~\ref{fig:milton_kine}) reproduce the same overall structure observed in the composite analysis. Meridional wind sensitivity remains concentrated between 850 and 500~hPa, while zonal wind sensitivity retains an upper-level maximum near 250~hPa. The close agreement between the individual-storm and composite results indicates that the learned kinematic representation is robust across storm-specific and basin-scale evaluations. Unlike the thermodynamic closure, however, the kinematic closure shows little systematic dependence on whether the storm is intensifying or weakening. This suggests that the neural representation of environmental shear behaves primarily as an external forcing whose vertical structure remains relatively stable across the storm lifecycle.

The contrast between the thermodynamic and kinematic saliency structures provides additional physical insight. The learned representation of $S$ remains nearly phase invariant, whereas the learned representation of $\chi$ becomes more strongly concentrated in the mid-troposphere during weakening, particularly through temperature sensitivity near 600~hPa. This behavior is consistent with the structure of the FAST equations, in which ventilation affects storm evolution through the moisture tendency term $\chi S m$. Because the effect of ventilation depends on the storm's evolving inner-core moisture state, thermodynamic susceptibility is expected to vary more strongly with storm phase than the external shear forcing itself. FAST-ML therefore appears to learn a physically plausible separation between a comparatively stable kinematic driver and a more state-dependent thermodynamic sink.
Taken together, these interpretability results provide strong evidence that the FAST-ML learns physically meaningful representations of environmental ventilation. The dominant thermodynamic sensitivities emerge in the 500--600~hPa layer expected from mid-level entropy-deficit theory, while the kinematic sensitivities recover the lower-to-upper-tropospheric wind structure associated with environmental vertical shear. These patterns are absent in randomly initialized networks, robust across attribution methods, and consistent between composite and individual-storm analyses. Beyond validating the architecture of FAST-ML, the results demonstrate that embedding machine learning within a physically constrained dynamical model can produce interpretable environmental parameterizations that remain closely aligned with established TC theory.

\subsection{Synthesis}

Across all evaluation frameworks, FAST-ML consistently improves upon the baseline FAST model through a more adaptive representation of environmental ventilation. The framework reduces deterministic forecast errors, improves rapid-intensification detection, alleviates recurrent lifecycle forecast failures, and delivers competitive probabilistic forecast skill in both independent and cross-basin evaluations. Architectural ablation experiments demonstrate that these improvements arise from the combined effects of physics-guided pretraining, the dual-stream closure architecture, and end-to-end optimization through the differentiable FAST solver. Selected case studies also show intensity predictions comparable to or closer to observations than FNV3 under the evaluated input configurations. The results indicate that the largest gains occur in situations where the original analytical ventilation closure performs poorly, suggesting that FAST-ML primarily improves the representation of storm--environment interaction rather than altering the underlying thermodynamic framework. The interpretability analyses further reinforce this conclusion, showing that the learned closure remains strongly aligned with established theories of mid-level ventilation and environmental shear despite being trained directly from high-dimensional atmospheric fields. Taken together, these findings suggest that forecast improvements arise primarily from more realistic estimation of unresolved environmental forcing, rather than from replacing or bypassing the governing physical dynamics. More broadly, the results indicate that physically constrained hybrid models can achieve forecast skill comparable to advanced data-driven systems while retaining transparency, physical consistency, and computational efficiency.

\section{Discussion and Conclusions}

This study developed FAST-ML, a hybrid TC intensity forecasting framework that embeds a neural environmental closure within the FAST reduced-order model. Rather than predicting storm intensity directly, the neural network diagnoses ventilation-related closure parameters from three-dimensional environmental fields, while storm evolution remains governed by the FAST thermodynamic equations. Across the evaluated experiments, FAST-ML reduces intensity forecast errors and improves rapid-intensification prediction relative to baseline FAST. Selected comparisons with FNV3 also show encouraging intensity predictions, subject to differences in input information and forecasting scope. The largest improvements occur in storms and forecast periods where the original analytical ventilation closure performs poorly, indicating that the primary contribution of FAST-ML is a more realistic representation of environmental forcing rather than modification of the underlying thermodynamic framework. A key outcome of this work is that forecast skill is improved without sacrificing physical interpretability. The learned closures remain consistent with established theories of ventilation and environmental shear despite being inferred directly from high-dimensional atmospheric fields. These results suggest that machine learning can be used to improve unresolved process representations while maintaining physically meaningful system behavior \cite{schneider2017,rasp2018,willard2022}. The framework also highlights the benefits of treating TC intensity forecasting as a specialized modeling problem. By focusing exclusively on intensity evolution and environmental forcing, FAST-ML can be coupled with externally generated track forecasts while remaining computationally efficient. More broadly, this work demonstrates the feasibility of replacing rigid heuristic parameterizations with machine-learning-based closures trained through end-to-end differentiable optimization, allowing observational data to constrain unresolved processes within a physically governed modeling framework \cite{raissi2019physics,karniadakis2021,willard2022}.

Several limitations should be acknowledged. First, this observation-calibrated differentiable paradigm is most suitable for systems whose governing equations are not strongly and deeply nonlinearly coupled. Second, FAST-ML inherits the structural assumptions of the FAST model, including its idealized axisymmetric representation of the TC vortex and simplified treatment of unresolved inner-core processes. Third, the latent moisture variable $m$ is not directly observable and therefore cannot be independently validated. Fourth, ensemble forecasts remain under-dispersive, indicating the need for improved probabilistic calibration. This likely reflects the strong deterministic constraints imposed by the underlying physical model and the reliance on externally generated ensemble perturbations. Finally, although the zero-shot Eastern Pacific experiments demonstrate encouraging transferability, broader evaluation across additional TC basins is required to assess global generalization.
Furthermore, while our current differentiable optimization focuses primarily on environmental ventilation parameters, the framework naturally extends to jointly optimizing other physical terms in the governing equations. For instance, the drag coefficient ($C_d$) could be dynamically constrained by surface friction conditions, and ocean coupling ($\alpha$) could be refined by directly incorporating subsurface oceanic observations. These potential extensions highlight the broader capacity of this differentiable paradigm to ingest heterogeneous observational datasets, paving the way for a deeper, data-driven resolution of the complex physical processes governing hurricane evolution.

Comparisons with FNV3 require care because the systems differ in both forecasting scope and input information. FNV3 predicts global atmospheric evolution, whereas FAST-ML predicts intensity conditional on externally supplied tracks and environmental fields. The ERA5-driven experiments are diagnostic, including those that use FNV3 track ensembles; shared tracks do not equalize environmental inputs. The GEFS-driven case studies demonstrate the feasibility of using forecast inputs, but do not establish general operational superiority. These comparisons therefore illustrate the potential complementary value of a specialized intensity model within a broader forecasting system.

Despite these limitations, the results demonstrate the value of embedding machine learning within reduced-order physical models to augment, rather than replace, established physical theory. The differentiable coupling between learned environmental closures and governing dynamical equations provides a practical pathway for improving unresolved parameterizations while retaining transparency, physical consistency, and computational efficiency. Although developed for TC intensity forecasting, the underlying framework is broadly applicable to other reduced-order Earth-system models in which unresolved processes remain a dominant source of uncertainty, including applications in hydrology, ocean dynamics, convection parameterization, and climate extremes \cite{rasp2018,willard2022,smithThorpe2026}. In summary, FAST-ML demonstrates that hybrid physics--machine learning frameworks can improve TC intensity prediction by learning unresolved environmental forcing while preserving physically based system evolution. More broadly, the study illustrates how machine learning can be integrated within reduced-order Earth-system models to improve predictive skill while maintaining interpretability and computational efficiency.

\acknowledgments

This work was supported by the Laboratory Directed Research and Development program at Sandia National Laboratories, a multimission laboratory managed and operated by National Technology and Engineering Solutions of Sandia, LLC, a wholly owned subsidiary of Honeywell International, Inc., for the U.S. Department of Energy's National Nuclear Security Administration under contract DE-NA-0003525.

This research used resources of the National Energy Research Scientific Computing Center (NERSC), a Department of Energy Office of Science UserFacility using NERSC award BER-ERCAP0037796.

This written work is partially authored by an employee of NTESS. The employee, not NTESS, owns the right, title, and interest in and to the written work and is responsible for its contents. Any subjective views or opinions that might be expressed in the written work do not necessarily
represent the views of the U.S. Government. The publisher acknowledges that the U.S. Government retains a non-exclusive, paid-up, irrevocable,
world-wide license to publish or reproduce the published form of this written work or allow others to do so, for U.S. Government purposes. The DOE will provide public access to results of federally sponsored research
in accordance with the DOE Public Access Plan. This paper describes objective technical results and analysis. Any subjective views or opinions that might be expressed in the paper do not necessarily represent the views of the U.S. Department of Energy or the United States Government.

\section*{Conflict of Interest}
The authors declare no conflicts of interest relevant to this study.
\section*{Data Availability Statement}

All datasets used in this study are publicly accessible. 
ERA5 atmospheric and surface reanalysis data are available through the Copernicus 
Climate Data Store at \url{https://doi.org/10.24381/cds.bd0915c6} and 
\url{https://doi.org/10.24381/cds.adbb2d47}. Tropical cyclone best-track positions 
and intensity observations are obtained from the NOAA International Best Track Archive 
for Climate Stewardship (IBTrACS), available at 
\url{https://doi.org/10.25921/82ty-9e16}. ECMWF Ensemble Prediction System (EPS) 
forecast data are obtained from the THORPEX Interactive Grand Global Ensemble 
(TIGGE) archive, publicly accessible through the ECMWF data portal at 
\url{https://apps.ecmwf.int/datasets/data/tigge/}. GEFS forecast fields used in the 
fully forecast-driven experiments are obtained from the NOAA/NCEP Global Ensemble 
Forecast System.

Baseline probabilistic forecasts from Google DeepMind's FNV3 are obtained from the 
Google DeepMind WeatherLab platform, publicly available at 
\url{https://deepmind.google.com/science/weatherlab}.

The complete source code used for model evaluation, and analysis is permanently archived on Zenodo at 
\url{https://doi.org/10.5281/zenodo.22765052}. The publicly available FAST-ML repository, is actively maintained and updated on GitHub at
\url{https://github.com/Shijie-Xiao/FAST-ML}. 
\newpage
\bibliography{agusample}
\end{document}